\documentclass[a4paper,notitlepage,twocolumn,superscriptaddress,longbibliography]{revtex4-1} 
\pdfoutput=1

\usepackage[utf8]{inputenc}
\usepackage{braket}
\usepackage{amsmath}
\usepackage{mathtools}
\usepackage{amssymb}
\usepackage{amsfonts}
\usepackage{graphicx}
\usepackage{hyperref}
\usepackage{comment}
\usepackage{bbold}
\usepackage{color}

\def\be{\begin{equation}}
\def\ee{\end{equation}}
\def\bea{\begin{eqnarray}}
\def\eea{\end{eqnarray}}

\DeclareUnicodeCharacter{03B1}{\ensuremath{\alpha}}
\DeclareUnicodeCharacter{03C0}{\ensuremath{\pi}}
\DeclareUnicodeCharacter{03A8}{\ensuremath{\Psi}}
\DeclareUnicodeCharacter{03C8}{\ensuremath{\psi}}
\DeclareUnicodeCharacter{03B3}{\ensuremath{\gamma}}
\DeclareUnicodeCharacter{03B4}{\ensuremath{\delta}}
\DeclareUnicodeCharacter{03C3}{\ensuremath{\sigma}}
\DeclareUnicodeCharacter{03C7}{\ensuremath{\chi}}
\DeclareUnicodeCharacter{03A9}{\ensuremath{\Omega}}
\DeclareUnicodeCharacter{03C1}{\ensuremath{\rho}}
\DeclareUnicodeCharacter{27E8}{\langle}
\DeclareUnicodeCharacter{27E9}{\rangle}

\newcommand{\lbar}{{\bar\ell}}

\newcommand{\tr}{\operatorname{tr}}

\begin{document}

\title{Quantum chaos in the Brownian SYK model with large finite $N$:\\
OTOCs and tripartite information}


\author{Christoph Sünderhauf}
\affiliation{Max-Planck-Institut für Quantenoptik, Hans-Kopfermann-Str.~1, 85748 Garching, Germany}
\affiliation{Munich Center for Quantum Science and Technology, Schellingstraße 4, 80799 M\"unchen, Germany}
\author{Lorenzo Piroli}
\affiliation{Max-Planck-Institut für Quantenoptik, Hans-Kopfermann-Str.~1, 85748 Garching, Germany}
\affiliation{Munich Center for Quantum Science and Technology, Schellingstraße 4, 80799 M\"unchen, Germany}
\author{Xiao-Liang Qi}
\affiliation{Stanford Institute for Theoretical Physics, Stanford University, Stanford, CA 94305, USA}
\affiliation{Department of Physics, Stanford University, Stanford, CA 94305, USA}
\affiliation{Google, 100 Mayfield Ave, Mountain View, CA 94043, USA}
\author{Norbert Schuch}
\affiliation{Max-Planck-Institut für Quantenoptik, Hans-Kopfermann-Str.~1, 85748 Garching, Germany}
\affiliation{Munich Center for Quantum Science and Technology, Schellingstraße 4, 80799 M\"unchen, Germany}
\author{J.~Ignacio Cirac}
\affiliation{Max-Planck-Institut für Quantenoptik, Hans-Kopfermann-Str.~1, 85748 Garching, Germany}
\affiliation{Munich Center for Quantum Science and Technology, Schellingstraße 4, 80799 M\"unchen, Germany}

\begin{abstract}

We consider the Brownian SYK model of $N$ interacting Majorana fermions, with random couplings that are taken to vary independently at each time.  We study the out-of-time-ordered correlators (OTOCs) of arbitrary observables and the R\'enyi-$2$ tripartite information of the unitary evolution operator, which were proposed as diagnostic tools for quantum chaos and scrambling, respectively. We show that their averaged dynamics can be studied as a quench problem at imaginary times in a model of $N$ qudits, where the Hamiltonian displays site-permutational symmetry. By exploiting a description in terms of bosonic collective modes, we show that for the quantities of interest the dynamics takes place in a subspace of the effective Hilbert space whose dimension grows either linearly or quadratically with $N$, allowing us to perform numerically exact calculations up to $N = 10^6$. We analyze in detail the interesting features of the OTOCs, including their dependence on the chosen observables, and of the tripartite information. We observe explicitly the emergence of a scrambling time $t^\ast\sim \ln N$ controlling the onset of both chaotic and scrambling behavior, after which we characterize the exponential decay of the quantities of interest to the corresponding Haar scrambled values. 

\end{abstract}

\maketitle

\tableofcontents

\section{Introduction}
\label{sec:intro}

The study of many body quantum chaos is currently experiencing a golden age, also due to its implications on important aspects in many-body physics such as the thermalization~\cite{	Deut91,Sred94} of isolated systems \cite{RiDO08,HoQi16, DKPR16, LeQi18,	HaMa13,LiSu14}, or the scrambling of quantum information \cite{HQRY16,LFSL19,BeGL19}. In fact, the field has already enjoyed intense research activity more than thirty years ago~\cite{Gutzwiller_book,stockmann_book}, when the relations between chaotic many-body systems and random matrix theory were first explored. Recently, a renewed interest came from the study of black hole physics and concepts such as scrambling of quantum information, and computational complexity~\cite{HaPr07,LSHO13, SeSu08,ShSt14, 	KitaevTalk2,Suss16,BRSS16}.

An important milestone in the recent literature is the Sachdev-Ye-Kitaev model~\cite{SaYe93,Kitaev_talk}, which was originally proposed by Sachdev and Ye as a model of strongly correlated electron systems, and generalized by Kitaev in 2015 who pointed out its connection to holographic duality. This model describes $N$ Majorana fermions or complex fermions with random all-to-all interactions. In the work we will focus on the Majorana fermion version. The SYK model has already drawn enormous attention from different communities, ranging from quantum gravity \cite{MaSt16,MaSY16} to condensed-matter and many-body physics \cite{PoRo16,BaAK16,DFGG17,KlTa17,GuQS17,SoJB17,CWBS18}, due to the concomitance of several unique features. Among these, the model has been shown to be maximally chaotic and yet amenable to exact analysis in the large-$N$ limit \cite{Kitaev_talk, PoRo16,MaSS16, BaAK16,BaAK17}, making it an ideal playground for the study of chaos and scrambling of quantum information. 

In the same years, the effort to better characterize quantum chaos led to the systematic development of reliable indicators for its diagnosis. In particular, out-of-time-ordered correlation (OTOC) functions, historically introduced in the context of disordered superconductors \cite{LaOv69}, were naturally selected as ideal probes to detect the ``scrambling'' of local observables \cite{ShSt14_bh, ShSt14, Kitaev_talk, RoSS15,RoSS18}, namely the spreading of their spatial support in the operator basis. It is important to mention that these ideas had far reaching ramifications, motivating the study of OTOCs also in many-body systems with short-range interactions \cite{RoSw16,TsWU17, AlFI16,SwCh17,Yung17,PaSa17,KuGP17,GuQS17,DoMo17,LiMo18,LiMo18,SKMK18,NIDS19,McNK19, 	HuBZ19,CLBS19} and in spatially local  ``quantum unitary circuits''~\cite{NRVH17,SPHS18, NaVH18,KRPS18,ChDC18,RaPK18,KhVH18,KoLP18,BeKP18,Hunt19,GuHu19,ZhNa19,FCDC19}, which provide minimally structured models for chaotic quantum dynamics. In fact, related studies on information scrambling in a class of random, and in general non-local, circuits (the so-called approximate $t$-designs) were already carried out within quantum information theory~\cite{EmLL05,EWSL03,DaOP07,GrAE07,OlDP07,Znid07,Znid08,ArBr08,HaLo09,BrVi10,DiJo11,BrFa12,BrHH16,NHKW17,SoCBQA19}, where the latter were used to provide rapid approximations to random unitary operators. Finally, we note that OTOCs were also shown to be directly related to the growth of the operator size, {\it i.e.} the size of its support~\cite{RoSS18,QiSt18}.

So far, computations of OTOCs in the SYK model have been carried out through field-theoretical approaches in the large-$N$ limit \cite{Kitaev_talk,MaSt16, PoRo16}. On the other hand, despite the many works devoted to this topic, results for finite values of $N$ are difficult to obtain, and remain scarce \cite{GaVe16,GaVe17, GaJV18}. This is also true for numerical computations: the exponential growth of the Hilbert space dimension, and the presence of disorder averages yield strong limitations on the sizes of the systems that can be simulated~\cite{FuSa16, RoSS18, CGHP17,GuMV18}.  Still, it would be highly desirable to develop a systematic approach to investigate the properties of the SYK model at finite $N$, even numerically. Indeed, not only would this allow for inspection of finite-size corrections to the large-$N$ results, but also to compute quantities beyond multi-point correlation functions, for which field-theoretical approaches might be difficult to apply. A notable example is given by the (negative) tripartite information of the evolution operator introduced in Ref.~\cite{HQRY16} in the context of unitary circuits. This was suggested as a valuable tool to quantify the scrambling power of a quantum channel, namely its ability to delocalize information provided as an input. We note that, so far, this quantity has been computed only numerically for small system sizes~\cite{HQRY16,SPMK18} (see also Refs.~\cite{IySa18,PRZI18,SeML18}, where the tripartite information of given states, and not of the channel, was studied).

Motivated by the above picture, we consider a simpler, but closely related, \emph{Brownian} SYK model, and address the problem of its exact analysis at finite sizes. The model was introduced in Ref.~\cite{SaSS18}, and differs from the traditional SYK in that the random couplings are chosen to vary independently at each time. The simplification arising in this case is similar to the one we have in unitary circuits by choosing random gates independently in each layer. Experience from the latter framework suggests that the main features of the chaotic dynamics remain qualitatively unaltered by introducing an additional time-dependence to the spatial disorder,  except that random circuits and Brownian models behave like infinite-temperature systems since they do not display energy conservation.

In this work, we focus on the development of a systematic approach to the chaotic dynamics in the Brownian SYK model, which could also be applied, more generally, to other time-dependent, disordered Hamiltonians with infinite-range interactions. In particular, we aim to compute OTOCs of arbitrary local observables, and other dynamical quantities which can be extracted from disordered averages involving up to four unitary evolution operators. These include a R\'enyi-$2$ version of the tripartite information introduced in \cite{HQRY16}, which has been shown to encode information about all possible OTOCs~ \cite{HQRY16}. 

As a main result of our work, we show that the averaged dynamics of the OTOCs and of the R\'enyi tripartite information can be studied as a quench problem at imaginary times in a model of $N$ qudits, where the Hamiltonian displays full site-permutational symmetry. We analyze this problem by means of a description in terms of bosonic collective modes, and prove that for the quantities of interest the dynamics takes place in a subspace of the Hilbert space whose dimension grows either linearly or quadratically with $N$. This allows us to perform numerically exact calculations up to one million particles, and, consequently, analyze in great detail the behavior of OTOCs and of the R\'enyi tripartite information, highlighting their most interesting features. While some of our results depend on simplifications arising in the special case of the SYK model, we expect that suitable generalizations of our method could be successfully applied also to the study of other disordered time-dependent Hamiltonians with all-to-all interactions.

It is useful to compare our method with that of existing studies, as some of the ideas used in our work are related to other approaches in the literature. First, Ref.~\cite{SaSS18} proposed the Brownian SYK model as a simplified version of the original SYK, and mainly focused on the computation of the spectral form factor~\cite{Haake_book}. For this specific quantity, it was shown that in the Brownian SYK model an exact solution could be achieved, by means of an elementary mapping to a classical partition function. Our results on OTOCs and tripartite information cannot be obtained using the same approach.

Next, we discuss Refs.~\cite{Znid07,Znid08,BrVi10}, where a class of random quantum circuits was considered, in which at each layer a single unitary gate is applied to a pair of qudits randomly chosen. There, it was shown that the moments of the evolution operator associated with a time step could be mapped onto a permutational invariant Hamiltonian which generalizes the Lipkin-Meshkov-Glick model~\cite{RiVM08}. Even though the idea underlying our method is similar, both our mapping and the quantities studied in this paper are different.

We also note that the computation of OTOCs in models with a continuous-time evolution in the presence of Brownian disorder and infinite-range interactions have been already addressed in \cite{ShSt15,ZhCh19} (see also \cite{ChZh18,XuSw18}). The system studied in these works, consisting of $N$ qudits driven by an Hamiltonian which is bilinear in the Pauli operators, was introduced as a chaotic toy model in Ref.~\cite{LSHO13}, where its scrambling time was first estimated to be logarithmic in $N$ (see also Ref.~\cite{GHST18}, where the spectral form factor was analyzed for the same system). The approach of \cite{ShSt15,ZhCh19} relies on the derivation, based on an application of It\^{o} calculus~\cite{Parthasarathy_book}, of a system of differential equations for the OTOCs of interest. Solving the latter, numerical results were given in Ref.~\cite{LSHO13} for sizes comparable to those that can be reached with our method, while an analytical solution was found in ~\cite{ZhCh19} for a particular average of OTOCs. As we will see, our approach differs from that of \cite{ShSt15,ZhCh19}, as we tackle directly the computation of the averaged moments of the evolution operator. This allows us to use the same formalism to also analyze the tripartite information discussed above, which was not addressed in these studies. Finally, we note that rigorous results, relevant to the present paper, for the scrambling properties of continuous-time evolution generated by random Hamiltonians were recently presented in Refs.~\cite{BaBK17,OBKB17}. 

The organization of the rest of this paper is as follows. In Sec.~\ref{sec:setup} we introduce the Brownian SYK model and the quantities which will be investigated in this work. We proceed to present the key features of our method in Sec.~\ref{sec:methods}, while our physical results are reported in Sec.~\ref{sec:results}. The most technical aspects of our study are consigned to Sec.~\ref{sec:technical_derivation} and to a few appendices. Finally, our conclusions are discussed in Sec.~\ref{sec:conclusions}.

\section{The model and the chaos quantifiers}
\label{sec:setup}

\begin{figure}
	\includegraphics[scale=0.8]{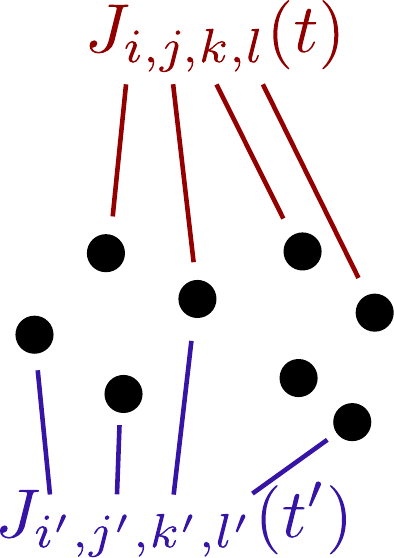}
	\caption{Pictorial representation of the Brownian SYK model described by the Hamiltonian~\eqref{eq:q_hamiltonian}, for $q=4$. At each time, all the Majorana fermions are coupled together within clusters of $q$ particles, with time-dependent random interactions.}
	\label{fig:SYK_model}
\end{figure}

The object of study of this work will be the Brownian SYK model, describing a set of $N$ Majorana fermions with $q$-local, all-to-all random interactions, cf. Fig.~\ref{fig:SYK_model}. It is defined on a Hilbert space $\mathcal{H}_N$ of dimension $D=2^{N/2}$, with $N$ operators $\psi_j$ acting on $\mathcal{H}_N$. They are the representation of standard Majorana fermions, and thus satisfy $\{\psi_j,\psi_k\}=2\delta_{j,k}$ and $\psi_j^{\dagger}=\psi_j$ (the quantities of interest in this work will not depend on the representation chosen for the $N$ Majorana fermions). Its time-dependent Hamiltonian reads
\be
H_{\rm SYK}(t)=i^{q/2}\sum_{1\leq i_1<i_2<\ldots <i_q\leq N}J_{i_1,\ldots \,, i_q}(t)\psi_{i_1}\psi_{i_2}\ldots \psi_{i_q}\,.
\label{eq:q_hamiltonian}
\ee
Here, the couplings $J_{i_1,\ldots\,, i_q}(t)$ are random variables, which we assume to be Gaussian distributed with vanishing mean and variance
\be
\overline{ J_{i_{1} \ldots i_{q}}(t) J_{i_{1}^{\prime} \ldots i_{q}^{\prime}}\left(t^{\prime}\right)}=\delta_{i_{1} i_{1}^{\prime}} \cdots \delta_{i_{q} i_{q}^{\prime}} \delta\left(t-t^{\prime}\right) \sigma_J \frac{(q-1) !}{N^{q-1}}\,,
\label{eq:sigma_J}
\ee
where we denoted by $\overline{[\ldots]}$ the average over disorder realizations. While our method could be applied for arbitrary integer values of $q$,  we will focus for concreteness on the case $q=4$. Furthermore, we will choose the constant $\sigma_J$ in such a way that
\be
\overline{J_{i_{1} \ldots\,, i_{4}}(t) J_{i_{1}^{\prime} \ldots \,, i_{4}^{\prime}}\left(t^{\prime}\right)}=\delta_{i_{1} i_{1}^{\prime}} \cdots \delta_{i_{4} i_{4}^{\prime}} \delta\left(t-t^{\prime}\right) \frac{1}{N^3}\,.
\label{eq:time_correlations}
\ee 
In comparison, the original SYK Hamiltonian shares the same form of \eqref{eq:q_hamiltonian}, but with time-independent couplings.
In appendix~\ref{sec:non-interacting} we additionally discuss the case $q=2$, which lacks chaotic behavior as each disorder realization is non-interacting.


\subsection{The OTOCs and the operator spreading}

As we have already discussed in Sec.~\ref{sec:intro}, we will be mainly interested in two quantifiers of quantum chaos and scrambling. The first one is given by OTOCs of local observables: explicitly, given two operators $\mathcal{O}$,  $\mathcal{O}^\prime$, we define their OTOC on a state $\rho$ as
\be
\mathcal{F}_{\mathcal{O},\mathcal{O}^\prime}(t)={\rm tr}\left[\rho \mathcal{O}(t)\mathcal{O}^\prime(0)\mathcal{O}(t)\mathcal{O}^\prime(0)\right]\,,
\label{eq:OTOCs}
\ee
where $\mathcal{O}(t)=U^{\dagger}(t)\mathcal{O}U(t)$, and $U(t)$ is the unitary evolution operator. In this work we will choose the infinite-temperature Gibbs density matrix $\rho=\mathbb{1}/2^{N/2}$, which represents a stationary state for the time-dependent Hamiltonian~\eqref{eq:q_hamiltonian}. 

Importantly, we recall that the OTOC~\eqref{eq:OTOCs} can be related to an intuitive notion of the spreading of localized operators under unitary evolution. To this end, we choose for simplicity $\mathcal{O}=\psi_j$, $\mathcal{O}^\prime=\psi_k$ with $j\neq k$, and consider the quantity
\be
\mathcal{C}(t)=\frac{1}{2}{\rm tr}\left[\rho\left(\left\{\psi_j(t),\psi_k(0)\right\}\right)^{\dagger}\left(\left\{\psi_j(t),\psi_k(0)\right\}\right)\right]\,,
\label{eq:anti_comm}
\ee
which measures the magnitude of the anticommutator between $\psi_j(t)$ and $\psi_k(0)$. At time $t=0$, one simply has $\mathcal{C}(t)=0$. On the other hand, as time increases, the spacial support of $\psi_j(t)$ will also increase; namely $\psi_j(t)$ will evolve into a complicated sum of strings of Majorana operators. Then, we see that deviations of $\mathcal{C}(t)$ from zero signal that the support of $\psi_j(t)$ has grown to include site $k$. Accordingly, $\mathcal{C}(t)$ can be understood as a measure of the spatial spreading of the local operator $\psi_j(t)$. The connection between the latter and OTOCs is finally established by the simple relation
\be
\mathcal{C}(t)=1+{\rm Re}\left[{\rm tr}\left(\rho \psi_j(t)\psi_k(0) \psi_j(t)\psi_k(0)\right)\right]\,.
\label{eq:relation_OTOC_commutator}
\ee
In conclusion, the discussion above allows one to view the  OTOCs as a measure of chaos: chaotic dynamics corresponds to OTOCs that vanish sufficiently rapidly with time. On the other hand, for a non-chaotic Hamiltonian one expects information to spread coherently: for large system sizes this results in either a slow decay or a non-vanishing asymptotics of OTOCs \cite{LiMo18}, while for small ones this causes revivals, consisting in OTOCs frequently returning close to their original value~\cite{GoYD19}.

\subsection{Diagnostic of scrambling: the tripartite information in fermionic systems}
\label{sec:intro_tripartite}

\begin{figure}
	\includegraphics[width=8.cm]{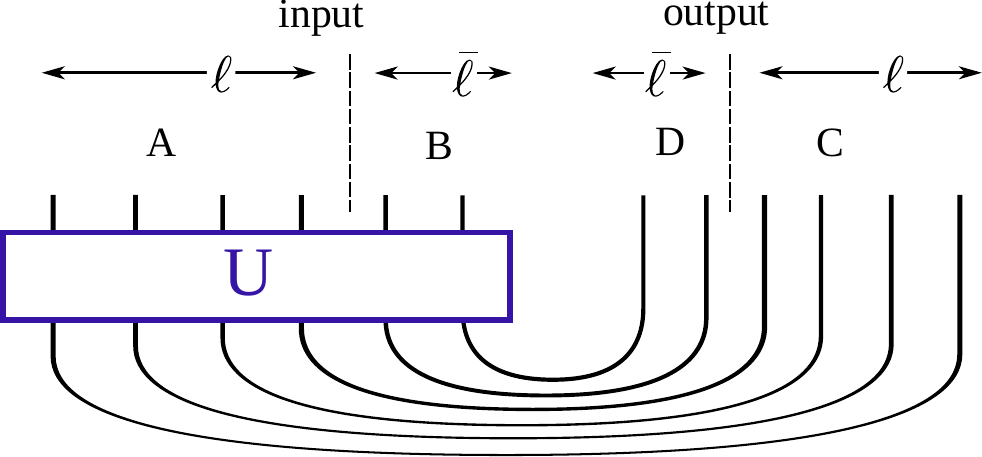}
	\caption{Pictorial representation of the state $\ket{U}\rangle$ defined in Eq.~\eqref{eq:state_U}. The operator $U$ is depicted as a box, while its legs correspond to the local Hilbert spaces $h_j$. The output legs are bent to create a state in a doubled Hilbert space $\mathcal{H}_N\otimes \mathcal{H}^\prime_N$. Each space is partitioned into two regions: $A$ and $B$ for the input space $\mathcal{H}_N$, and $C$ and $D$ for the output space $\mathcal{H}^\prime_N$.}
	\label{fig:entropy_definition}
\end{figure}

The OTOCs provide a physically clear definition of quantum chaos in terms of correlation functions between local operators. Other measures probing different features intuitively associated with chaos exist. Among these, the notion of scrambling of information, originally introduced in the study of black hole physics~\cite{HaPr07,SeSu08}, is particularly clear: a quantum system is a good scrambler if a localized perturbation in the initial state spreads over all its degrees of freedom, in such a way that it can no longer be detected by local measurements at large times. In this context, it is useful to think of the unitary evolution as a quantum channel, taking an initial state as the input, and returning the evolved state as the output. In this logic, it was proposed in Ref.~\cite{HQRY16} that the scrambling power of a channel could be conveniently measured by the tripartite information between bipartitions of its input and output, as we review in the following.

For simplicity, let us first consider a system of $N$ qudits, associated with a Hilbert space $\mathcal{H}_N=h_{1}\otimes \ldots \otimes  h_{N}$, where $h_j\simeq \mathbb{C}^{D}$, and a unitary operator $U:\mathcal{H}_N\to \mathcal{H}_N$. In order to study the scrambling properties of $U$, we wish to interpret it as a state in a suitable space. To this end, we introduce a copy of the original Hilbert space $\mathcal{H}^\prime_N$, and define the maximally entangled state $\ket{I}\in \mathcal{H}_N\otimes \mathcal{H}^\prime_N$ as
\be
\ket{I}=\frac{1}{D^{N/2}}\sum_{j=1}^{D^N}\ket{j}\otimes \ket{j'}\,,
\ee
where $\{\ket{j}\}_{j=1}^{D^N}$, $\{\ket{j^\prime}\}_{j^\prime=1}^{D^N}$ are orthonormal bases for $\mathcal{H}_N$ and $\mathcal{H}^\prime_N$, respectively. Note that we choose the basis such that $\ket{I}$ is a direct product of EPR pairs between qudits in the two systems, as is illustrated in Fig. \ref{fig:entropy_definition}. Then, the operator $U$ can be interpreted as a state in $\mathcal{H}_N\otimes \mathcal{H}_N^\prime$ through the Choi-Jamiolkowski mapping
\be
U\mapsto \ket{U}\rangle=\left(\mathbb{1}\otimes U\right) \ket{I}\,.
\label{eq:state_U}
\ee
Here the operator $U$ is depicted as a box, whose legs correspond to the local Hilbert spaces $h_j$; we see that one could intuitively think of the state $\ket{U}\rangle$ as obtained by ``bending'' the output legs, so as to treat input and output, associated with  $\mathcal{H}_N$ and $\mathcal{H}^\prime_N$ respectively, on an equal footing. It should be noted that the mapping from $U$ to $\ket{U}\rangle$ is not unique, as it depends on the choice of state $\ket{I}$. However, different $\ket{I}$ are related by a local unitary transformation, which does not affect the entropy-related quantities we discuss in the following.

Given $\ket{U}\rangle \in \mathcal{H}_N\otimes \mathcal{H}^\prime_N$, one can compute the entanglement entropy between different spatial regions in $\mathcal{H}_N$ and $\mathcal{H}^\prime_N$. We consider in particular
bipartitions of $\mathcal{H}_N$ and $\mathcal{H}^\prime_N$ into the complementary subsystems $A$, $B$ and $C$,$D$ respectively; in Fig.~\ref{fig:entropy_definition} a special choice for these regions is shown. Given a pair of bipartitions $(A,B)$ and $(C,D)$, we define the tripartite information as~\cite{HQRY16}
\be
I_{3}(A : C : D) = I(A : C)+I(A : D)-I(A : C D)\,,
\label{eq:def_tripartite}
\ee
where $CD$ denotes the union of the regions $C$ and $D$. Here $I(X:Y)$ is the mutual information between the regions $X$ and $Y$ 
\be
I(X : Y)=S_{X}+S_{Y}-S_{X Y}\,,
\ee
where $S_{X}$ is the von Neumann entropy of the reduced density matrix $\rho_{X}$. For instance, we have
\be
S_{A C}=-\operatorname{tr}\left[\rho_{A C} \ln \rho_{A C}\right]\,,
\ee
where $\rho_{A C}=\operatorname{tr}_{B D}[\rho]$.

The tripartite information in Eq.~\eqref{eq:def_tripartite} was suggested in Ref.~\cite{HQRY16} as a natural and convenient diagnostic for scrambling. In fact, as in the case of OTOCs, its underlying physical meaning is easy to grasp. From Eq.~\eqref{eq:def_tripartite}, we see that  $-I_3(A:C:D)$ quantifies the amount of information on the input region $A$ that can be recovered by global measurements in $C\cup D$, but can not be obtained by probing $C$ and $D$ individually. Recalling that $\mathcal{H}^\prime_N=C\cup D$ corresponds to the output, this is exactly a measure of scrambling: if $-I_3(A:C:D)$ is large, it means that the information localized in a subsystem $A$ of the input state can be recovered only by global measurements in the output state, and information has been scrambled. Accordingly, if for any bipartition of $\mathcal{H}_{N}$ and $\mathcal{H}^{\prime}_{N}$, $I_{3}(A:C:D)$ is negative with an absolute value close to the maximum possible value, the channel $U$ has large scrambling power. Finally, a close connection was established  in Ref.~\cite{HQRY16} between the tripartite information \eqref{eq:def_tripartite} and the OTOCs, which further corroborated the appeal of the former as a valuable diagnostic of scrambling and, more generally, of chaotic dynamics. This connection is reviewed in Appendix~\ref{sec:tripartite_and_otocs}, where we also discuss its generalization to the fermionic setting.

The above discussion is carried out in terms of qudits, whereas in our work we are interested in a fermionic system. At this point, one could employ a Jordan-Wigner representation of the Majorana operators in the Hamiltonian \eqref{eq:q_hamiltonian}, interpret the resulting evolution operator as a unitary channel acting on a system of $N/2$ qubits, and define the tripartite information for the latter according to the discussion above. However, given a correspondence between Majorana and Pauli operators via the Jordan-Wigner transformation, it is known that the reduced density matrix of disjoint intervals written in terms of the two is not the same, leading to different results for the corresponding von Neumann entanglement entropy~\cite{IgPe10,FaCa10}. In our case, we stress that the physical degrees of freedom are represented by the Majorana operators and, accordingly, the tripartite information in Eq.~\eqref{eq:def_tripartite} should be computed in terms of the latter. In this respect, we find it useful to discuss explicitly the generalization of the above construction for Majorana operators, without making direct reference to the tensor-product structure of the doubled Hilbert space associated with the input and output of the channel. 

\begin{figure}
	\includegraphics[width=8cm]{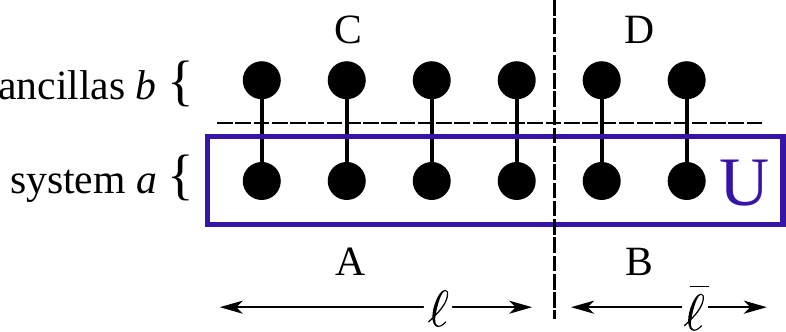}
	\caption{Pictorial representation for the state $\ket{I^{ab}}$ in Eq.~\eqref{eq:I_state_fermions}. The black bullets in the lower and upper rows represent the original and ``replica'' fermions $\psi^{a}_j$ and $\psi^{b}_j$, respectively. Each link is a maximally entangled pair, which corresponds to the vacuum for the complex Fermi operators $c_j=\psi^{a}_j-i\psi^b_j$. The evolution operator $U$, generated by the Hamiltonian \eqref{eq:q_hamiltonian}, is applied only to the original system.}
	\label{fig:entropy_majorana}
\end{figure}

As a first ingredient, we wish to interpret the evolution operator generated by the Hamiltonian~\eqref{eq:q_hamiltonian} as a state. To this end, we introduce a system of $2N$ Majorana operators $\psi^\alpha_j$, where $j=1\,,\ldots\,, N$, while $\alpha$ is an index labeling two different species which we denote by $a$ and $b$. The maximally entangled state $\ket{I^{ab}}$ is then defined as the vacuum state for the complex fermions $c_j=\psi_j^{a}-i\psi_j^{b}$~\cite{GuLQ17}, namely
\be
\left(\psi_j^{a}-i\psi_j^{b}\right)\ket{I^{ab}}=0\,,\qquad \forall j\,.
\label{eq:I_state_fermions}
\ee
The operator $U$ can now be interpreted as a state in the doubled system through the mapping
\be
U(t)\mapsto \ket{U(t)}\rangle=U^{a}(t) \ket{I^{ab}}\,.
\label{eq:state_U_fermions}
\ee
Here the superscript $a$ indicates that the Hamiltonian generating the unitary evolution operator $U^{a}(t)$ is written in terms of the fermions $\psi^{a}_j$. A pictorial representation of this construction is shown in Fig.~\ref{fig:entropy_majorana}. One can now proceed to compute the fermionic reduced density matrices for the evolved state $\ket{U(t)}\rangle$, and consequently the corresponding tripartite information as in Eq.~\eqref{eq:def_tripartite}. We refer the reader to Sec.~\ref{sec:technical_derivation} for further details.

Unfortunately, despite its great interest, the computation of the tripartite information \eqref{eq:def_tripartite} is a very difficult task, which so far has been carried out only numerically for qudit systems of small sizes~\cite{HoQi16,SPMK18}. For this reason, we study a simpler but closely related quantity, which is obtained from $I_3(A:C:D)$ by considering R\'enyi, rather than von Neumann, entropies. Specifically, we will compute the following R\'enyi-$2$ tripartite information
\be
I^{(2)}_{3}(A : C : D) = I^{(2)}(A : C)+I^{(2)}(A : D)-I^{(2)}(A : C D)\,,
\label{eq:def_tripartite_renyi}
\ee
where
\be
I^{(2)}(X : Y)=S^{(2)}_{X}+S^{(2)}_{Y}-S^{(2)}_{X Y}\,,
\ee
and
\be
S^{(2)}_{X}=-\ln\left[\overline{\operatorname{tr}\left(\rho^2_{X}\right)}\right]\,.
\label{eq:Renyi_2}
\ee
We note that, strictly speaking, $S^{(2)}_X$ is not the averaged R\'enyi entropy of order $2$, as the disorder average is taken inside the logarithm. However, Ref. \cite{HoQi16} showed that the OTOC for a pair of operators in $A$ and $C$ averaged over all operator choices is determined by ${\rm tr}\left(\rho_{AD}^2\right)$. Therefore the averaged purity $\overline{{\rm tr}\left(\rho_X^2\right)}$ is a meaningful physical quantity to consider. Also, for $N$ not too small, one expects the effect of fluctuations in the disorder to be small, so that $S^{(2)}_X$ remains a good approximation for the R\'enyi-$2$ entropy~\cite{HoQi16,KRPS18}. 

It is worth to notice that Eq.~\eqref{eq:def_tripartite_renyi} can be simplified in general. Indeed, it is easy to show~\cite{HQRY16}
\be
I^{(2)}_3(A:C:D)=\frac{N}{2}\ln(2)-S^{(2)}_{AC}-S^{(2)}_{AD}\,,
\label{eq:operative_formula}
\ee
where we used that the dimension of the Hilbert space associated with $N$ Majorana fermions is $D=2^{N/2}$. Eq.~\eqref{eq:operative_formula} tells us that, in order to obtain the tripartite information, it is sufficient to compute the entropies $S^{(2)}_{AC}$ and $S^{(2)}_{AD}$ between different regions of the input and the output.

We conclude this section by stressing that while the R\'enyi tripartite information \eqref{eq:def_tripartite_renyi} differs quantitatively from $I_{3}(A:C:D)$,  based on previous studies~\cite{GuLQ17}, we can still expect it to display the same qualitative features of the latter, and thus to be a suitable measure for scrambling.

\section{Exact approach from emergent permutational symmetry}
\label{sec:methods}

Having introduced the model and the quantities of interest, we proceed by presenting the general ideas of the method developed in this work. The physical results will be then discussed in Sec.~\ref{sec:results}, while we postpone the most technical details of our calculations to Sec.~\ref{sec:technical_derivation}. 

\subsection{Decomposing the dynamical problem}
\label{sec:decomposing the dynamical problem}

We will begin our discussion with the concrete problem of computing the OTOC~\eqref{eq:OTOCs}, which we rewrite as
\be
\mathcal{F}_{\mathcal{O},\mathcal{O}^\prime}(t) =\frac{1}{2^{N/2}} {\rm tr} \left\{\mathcal{O}U(t) \mathcal{O}^\prime U^\dagger(t) \mathcal{O}U(t)\mathcal{O}^\prime U^\dagger(t)\right\}\,.
\label{eq:general_OTOC}
\ee
We recall that the time-dependent, disordered Hamiltonian~\eqref{eq:q_hamiltonian} gives rise to a dynamics which can be interpreted as the continuous limit of the discrete process defined by 
\be
U(t)=e^{ -i \Delta t H_{\rm SYK}(t_n)}e^{- i \Delta t H_{\rm SYK}(t_{n-1})}\ldots e^{- i\Delta t  H_{\rm SYK}(t_1)}\,,
\label{eq:generator}
\ee
where $\Delta t=t/n$ and $t_{j}=j\Delta t$, while the delta function in Eq.~\eqref{eq:time_correlations} is regularized as 
\be
\overline{J_{i_{1} \ldots\,, i_{4}}(t_r) J_{i_{1}^{\prime} \ldots \,, i_{4}^{\prime}}\left(t_s\right)}=\delta_{i_{1} i_{1}^{\prime}} \cdots \delta_{i_{4} i_{4}^{\prime}}\frac{1}{N^3}\frac{\delta_{r,s}}{\Delta t}\,.
\label{eq:discrete_correlations}
\ee
In practice, one can work with the discrete form \eqref{eq:generator} of the evolution operator, and take the continuum limit at the end of the calculations.

\begin{figure}
	\includegraphics[width=8.5cm]{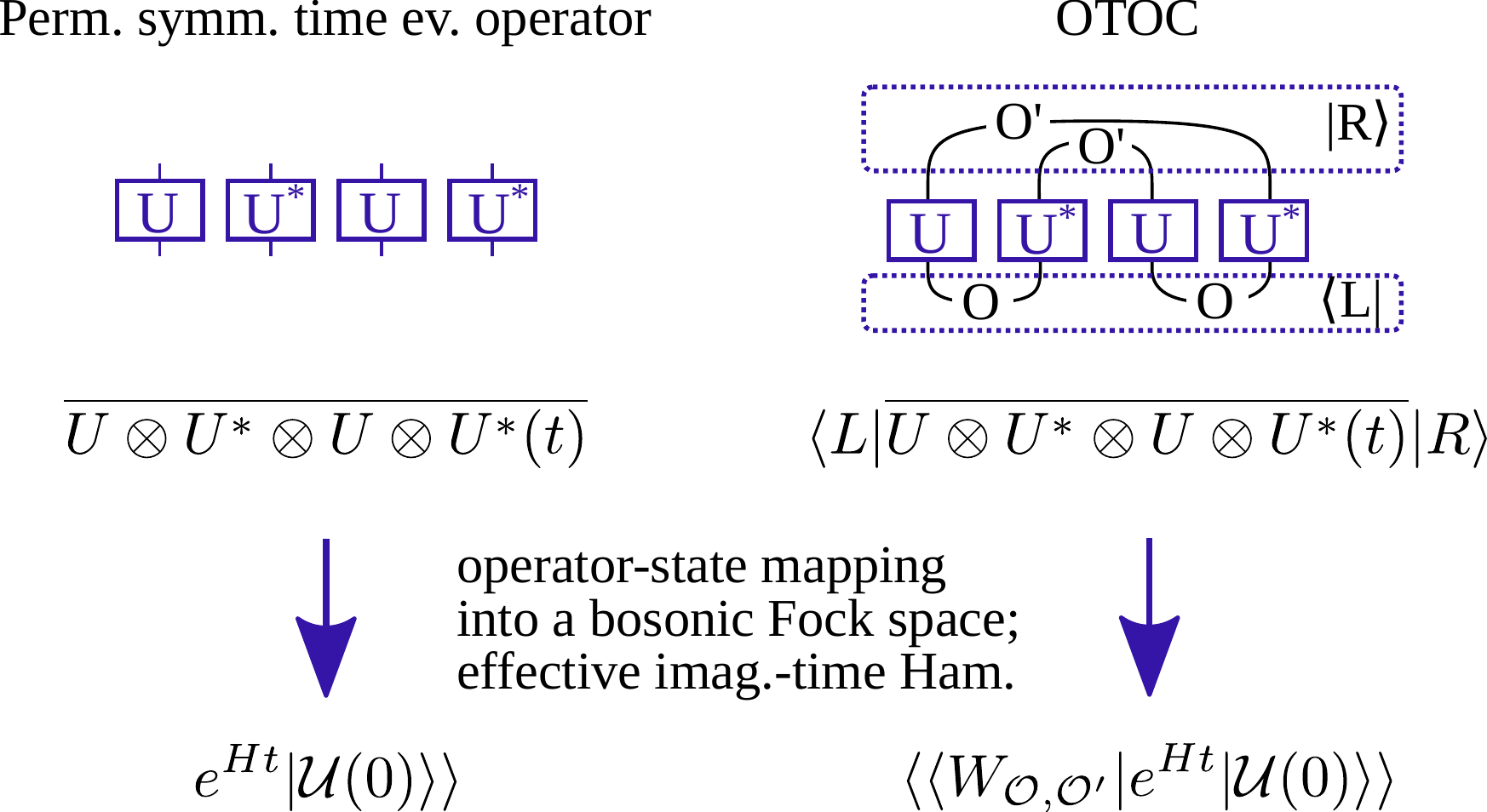}
	\caption{Schematic diagram summarizing our method. The dynamical information about the OTOC is uniquely encoded in the operator $\mathcal{U}(t) \equiv U\otimes U^*\otimes U \otimes U^*(t)$, while the observables $\mathcal{O},\mathcal{O'}$ define the ``right'' and ``left'' states $\ket{L}$, $\ket{R}$ (cf. Sec~\ref{sec:decomposing the dynamical problem}). Exploiting the emergent permutational symmetry, we can map $\overline{U\otimes U^*\otimes U\otimes U^*(t)}$ onto a state $\ket{\overline{\mathcal{U}}(t)}\rangle$ in a bosonic Fock space, in which the dynamics is governed by an effective imaginary-time Hamiltonian evolution (cf. Sec.~\ref{sec:generator_dynamics}). Finally, we express the matrix element of $\overline{\mathcal{U}}(t)$ with respect to $\ket{L}$ and $\ket{R}$ as the overlap between $\ket{\overline{\mathcal{U}}(t)}\rangle$ and an appropriate state $\langle\bra{W_{\mathcal{O},\mathcal{O'}}}$ (cf. Sec~\ref{sec:OTOC_INFO}). As a result, the entire computation of the OTOC can be performed very efficiently within a bosonic space, whose dimension grows linearly with $N$. The R\'enyi-$2$ entanglement entropies $S^{(2)}_{AC}$ and $S^{(2)}_{AD}$ are amenable to a similar treatment as the OTOCs.} 
	\label{fig:OTOC_pic}
\end{figure}

In order to compute $\mathcal{F}_{\mathcal{O},\mathcal{O}^\prime}(t)$, we first introduce a resolution of the identity between each pair of operators in ~\eqref{eq:general_OTOC}, yielding
\begin{align}
\mathcal{F}_{\mathcal{O},\mathcal{O}^\prime} &= \frac{1}{2^{N / 2}}\sum_{\substack{i, j, k, l \\ m, n, o, p }}\langle i |\mathcal{O} | j \rangle \langle j|U| k\rangle \left\langle k\left|\mathcal{O}^\prime\right| l\right\rangle \left \langle l\left|U^{\dagger}\right| m\right\rangle \nonumber\\
\times & \langle m |\mathcal{O}| n\rangle \langle n|U| o\rangle \left\langle o\left|\mathcal{O}^\prime\right| p\right\rangle\left\langle p\left|U^{\dagger}\right| i\right\rangle\,.
\label{eq:otoc_reinsertion}
\end{align}
Here $\{\ket{j}\}$ denotes a basis for the Hilbert space $\mathcal{H}_N$ [introduced before Eq.~\eqref{eq:q_hamiltonian}] on which the operators $\psi_j$ act. Rearranging the above sum, we obtain
\begin{align}
\mathcal{F}_{\mathcal{O},\mathcal{O}^\prime}= \bra{L} \left( U\otimes U^\ast\otimes U\otimes U^\ast \right) \ket{R} \,,
\label{eq:intermediate_1}
\end{align}
where
\begin{align}
\ket{L}=&\sum_{\substack{i, j, m, n}}\langle i | \mathcal{O}  | j \rangle  \langle m | \mathcal{O} | n\rangle\ket{j,m,n,i}\,,\\
\ket{R}=&\sum_{\substack{k, l o, p }} \left\langle k\left| \mathcal{O}^\prime \right| l\right\rangle \left\langle o\left| \mathcal{O}^\prime  \right| p\right\rangle\ket{k,l,o,p}\,.
\label{eq:L,R otocs}
\end{align}
Here $U^\ast(t)$ denotes the complex conjugate of $U(t)$ (which is well defined, once a basis $\{\ket{j}\}$ of $\mathcal{H}_N$ is given) and we introduced the vectors $\ket{i,j,k,l}=\ket{i}\otimes\ket{j}\otimes\ket{k}\otimes\ket{l}\in \mathcal{H}_N^{\otimes 4}$. According to Eq.~\eqref{eq:intermediate_1}, the dynamical information about the OTOC is uniquely encoded in the operator $\mathcal{U}(t)\equiv  U\otimes U^\ast\otimes U\otimes U^\ast $, while $\mathcal{O}$, $\mathcal{O}^\prime$ only affect the ``left'' and ``right'' states $\ket{L}$, $\ket{R}$, cf. Fig.~\ref{fig:OTOC_pic} .

From Eq.~\eqref{eq:generator}, we see immediately that $\mathcal{U}(t)$ is written in terms of the operators 
\begin{align}
\chi^{a}_j:=&\psi_j\otimes \mathbb{1}\otimes \mathbb{1}\otimes \mathbb{1}\,,\quad 
\chi^{b}_j:= \mathbb{1}\otimes \psi^\ast_j\otimes \mathbb{1}\otimes \mathbb{1}\,, \label{eq:chi_1}\\
\chi^{c}_j:=& \mathbb{1}\otimes \mathbb{1}\otimes \psi_j\otimes \mathbb{1}\,,\quad
\chi^{d}_j:=\mathbb{1}\otimes \mathbb{1}\otimes \mathbb{1}\otimes \psi_j^\ast\,,\label{eq:chi_2}
\end{align}
which provide a basis for all the operators in $\mathcal{H}^{\otimes 4}_N$. Note that, as we already stressed, $\psi_j$ is the representation of a Majorana fermion, and thus is an operator acting on $\mathcal{H}_N$, for which the tensor product is defined in the usual way. Due to the tensor-product structure of $\mathcal{H}_N^{\otimes 4}$, the operator $\chi_j^\alpha$ satisfy mixed commutation and anticommutation relations, namely $\left[\chi^{\alpha}_j,\chi^{\beta}_k\right]=0$ if $\alpha \neq \beta$, while $\left\{\chi^{\alpha}_j,\chi^{\alpha}_k\right\}=2\delta_{j,k}$. On the other hand, it is possible to introduce related operators in $\mathcal{H}_N^{\otimes 4}$ which are all anti-commuting with one another, realizing a truly fermionic algebra. We consider for concreteness the case $N\equiv 0\ ({\rm mod}\ 4)$ [if $N\equiv 2\ ({\rm mod}\ 4)$, one has a similar treatment], and introduce
\be
\mathcal{Q}^{\alpha}=\prod_{k=1}^N\chi^{\alpha}_k\,,\quad \alpha=a\,,b\,,c\,,d\,.
\ee
Then, we can define
\begin{align}
\psi^{a}_j=&i\mathcal{Q}^a\chi^{a}_j\,,\quad \psi^{b}_j=\mathcal{Q}^a\chi^{b}_j\,,\\
\psi^{c}_j=&i\mathcal{Q}^a\mathcal{Q}^b\mathcal{Q}^c\chi^{c}_j\,,\quad \psi^{d}_j=\mathcal{Q}^a\mathcal{Q}^b\mathcal{Q}^c\chi^{d}_j\,.
\end{align}
One can easily verify that $\{\psi^\alpha_j\}_{j,\alpha}$ satisfy fermionic anticommutation relations, namely $\{\psi_j^{\alpha},\psi_k^\beta\}=2\delta_{\alpha,\beta}\delta_{j,k}$, and that $\left(\psi^\alpha_j\right)^{\dagger}=\psi^\alpha_j$. Furthermore, since $\left(\mathcal{Q^{\alpha}}\right)^2=\mathbb{1}$, we have $\prod_{k=1}^M \chi_{j_k}^{\alpha}=\prod_{k=1}^M \psi_{j_k}^{\alpha}$ for any even integer $M$. Since the Hamiltonian \eqref{eq:q_hamiltonian} contains a sum of products of Majorana operators with an even number of particles, it is then straightforward to show
\be
\mathcal{U}(t)=U_+^a(t)U_-^b(t)U_+^c(t)U_-^d(t)\,,
\label{eq:generator_dynamics}
\ee
where
\be 
U^{\alpha}_\pm(t)=e^{\mp i \Delta t H_{\rm SYK}^{\alpha}(t_n)}e^{\mp i \Delta t H_{\rm SYK}^{\alpha}(t_{n-1})}\ldots e^{\mp i\Delta t  H_{\rm SYK}^{\alpha}(t_1)}\,,
\label{eq:signed_generator}
\ee
while $H_{\rm SYK}^{\alpha}$ is the Hamiltonian \eqref{eq:q_hamiltonian} written in terms of the fermions $\psi^{\alpha}_j$. We see that $\mathcal{U}(t)$ can be viewed as an evolution operator on the space of four ``replica'' Majorana fermions $\psi^{\alpha}_j$, labeled by $\alpha=a$, $b$, $c$, $d$. Eq.~\eqref{eq:generator_dynamics} represents the starting point for our subsequent calculations.

The above discussion allows us to decompose the problem of computing the OTOC \eqref{eq:general_OTOC} into two logically separated steps:
\begin{itemize}
	\item compute the disorder average of the generator of the dynamics $\mathcal{U}(t)$, defined in Eq.~\eqref{eq:generator_dynamics} (cf.~Sec.~\ref{sec:generator_dynamics});
	\item given the operator $\overline{\mathcal{U}}(t)$, evaluate the matrix element $\braket{L|\overline{\mathcal{U}}(t)| R}$, where $\ket{L}$, $\ket{R}$ were introduced in Eq.~\eqref{eq:intermediate_1} and pictorially represented in Fig.~\ref{fig:OTOC_pic} (cf.~Sec.~\ref{sec:OTOC_INFO}).
\end{itemize}
Importantly, it is possible to show that the same procedure can be employed for the R\'enyi-$2$ tripartite information \eqref{eq:def_tripartite_renyi}: one can express also this quantity in the form of a matrix element $\braket{L|\overline{\mathcal{U}}(t)|R}$, for an appropriate choice of the vectors $\ket{L}$ and $\ket{R}$, cf. Sec.~\ref{sec:OTOC_INFO}. We will address the two points above separately in the following subsections, for both the OTOCs and the tripartite information, providing a complete overview of the approach developed in this work.

\subsection{The generator of the dynamics:\\
	 mapping to a bosonic system}
 \label{sec:generator_dynamics}

We start by addressing the computation of the average evolution operator defined in Eq.~\eqref{eq:generator_dynamics}. Using that even numbers of different Majorana operators commute, and that one can factor disorder averages at different times, we note that  Eqs.~\eqref{eq:generator_dynamics}, \eqref{eq:signed_generator} imply
\begin{align}
\overline{\mathcal{U}}(t_n)=&\overline{e^{-i\Delta t H^{a}(t_n)}e^{i\Delta t H^{b}(t_n)}e^{-i\Delta t H^{c}(t_n)}e^{i\Delta t H^{d}(t_n)}}\nonumber\\
&\times \overline{\mathcal{U}}(t_{n-1})\,.
\end{align}
This allows us to write down a linear differential equation for $\overline{\mathcal{U}}(t)$, as follows. 

First, from Eq.~\eqref{eq:discrete_correlations}, we see that, in order to expand  the first line at the first order in $\Delta t$, each exponential factor has to be expanded up to the \emph{second} order. By doing this, and carefully taking into account the correlations between the couplings, one obtains an equation of the form
\be
\overline{\mathcal{U}}(t_n)=\overline{\mathcal{U}}(t_{n-1})+ L\overline{\mathcal{U}}(t_{n-1})\Delta t+O(\Delta t^2)\,,
\ee
namely, taking the limit $\Delta\to 0$
\be
\frac{{\rm d}}{{\rm d}t}\overline{\mathcal{U}}(t)= L \overline{\mathcal{U}}(t)\,,
\label{eq:diff_equation}
\ee
where
\begin{align}
L=\frac{1}{N^3}\Bigg[-2\binom{N}{4} +\sum_{\substack{\alpha,\beta =a,b,c,d \\\alpha<\beta}}(-1)^{\gamma_{\alpha,\beta}}&\nonumber\\
\times \left.\sum_{i_1<i_2<i_3<i_4}\left(\psi^{\alpha}_{i_1}\psi^{\beta}_{i_1}\right)\left(\psi^{\alpha}_{i_2}\psi^{\beta}_{i_2}\right)\left(\psi^{\alpha}_{i_3}\psi^{\beta}_{i_3}\right)\right.&\left(\psi^{\alpha}_{i_4}\psi^{\beta}_{i_4}\right)\Bigg]\,.
\label{eq:replica_hamiltonian}
\end{align}
Here, the indexes $a,b,c,d$ are ordered as $a<b<c<d$, while we introduced
\be
(-1)^{\gamma_{\alpha,\beta}}=
\begin{cases}
	1 & (\alpha,\beta)=(a,b), (a,d), (b,c), (c, d)\,,\\
	-1 & (\alpha,\beta)=(a,c), (b,d)\,.
\end{cases}
\label{eq:gamma_sign}
\ee
We note that, since the disorder average has been already taken, the operator $L$ is time- and disorder-independent. Eq.~\eqref{eq:diff_equation} can thus be seen as a Schrodinger equation (at imaginary times) for $\overline{\mathcal{U}}(t)$ in the space ${\rm End}(\mathcal{H}_N^{\otimes 4})$ of the linear endomorphisms acting on $\mathcal{H}_N^{\otimes 4}$, where the left matrix multiplication by $L$ is interpreted as a superoperator. In the following, it will be useful to denote by $\ket{\mathcal{O}}\rangle$ the state in ${\rm End}(\mathcal{H}_N^{\otimes 4})$ associated with the operator $\mathcal{O}$.

In order to proceed, we note that at every time $t$, the operator $\overline{\mathcal{U}}(t)$ can be written as a linear superposition of operators of the form
\be
\mathcal{O}_{1}^{\alpha_{1}}\mathcal{O}_{2}^{\alpha_{2}}\ldots \mathcal{O}_{N}^{\alpha_{N}}\,,
\label{eq:operator_product}
\ee
where $\mathcal{O}_{j}^{\alpha_{j}}$ is chosen within the set of operators
\begin{align}
I_{j}=\{\mathbb{1},\psi^{a}_{j}\psi^{b}_{j}\,, \psi^{a}_{j}\psi^{c}_{j}, \psi^{a}_{j}\psi^{d}_{j} \}&\nonumber\\
\cup \{\psi^{b}_{j}\psi^{c}_{j}\,,\psi^{b}_{j}\psi^{d}_{j}\,,&\psi^{c}_{j}\psi^{d}_{j}, \psi^{a}_{j}\psi^{b}_{j}\psi^{c}_{j}\psi^{d}_{j}\}\,.
\label{eq:I_set}
\end{align}
Indeed, due to the anticommutation relations of the Majorana operators and the form of the Hamiltonian $H$, it is easy to see that $\overline{\mathcal{U}}(t)$ can not contain terms with an odd number of fermions at site  $\psi^\alpha_j$. Hence, the dynamics of $\ket{\overline{\mathcal{U}}(t)}\rangle$ takes places in the Hilbert space generated by the vectors
\be
\ket{\alpha_1\ldots \alpha_{N}}:=\ket{\mathcal{O}_{1}^{\alpha_{1}}\mathcal{O}_{2}^{\alpha_{2}}\ldots \mathcal{O}_{N}^{\alpha_{N}}}\rangle\,.
\label{eq:mapping}
\ee
Here, $\alpha_j\in\{1,ab,ac,ad,bc,bd,cd,abcd\}$, with the convention $\mathcal{O}_j^{1}=\mathbb{1}$, $\mathcal{O}_j^{ab}=\psi^{a}_{j}\psi^{b}_{j}$, $\ldots$, $\mathcal{O}_j^{abcd}=\psi^{a}_{j}\psi^{b}_{j}\psi^{c}_{j}\psi^{d}_{j}$, i.e the ordered set $\{\mathcal{O}_j^{\alpha}\}_{\alpha=1}^{abcd}$ coincides with $I_{j}$ in~Eq.~\eqref{eq:I_set}.

Eq.~\eqref{eq:mapping} defines the previously announced mapping to a system of $N$ qudits, as one can interpret 
\be
\ket{\alpha_1 \ldots \alpha_{N}}=\ket{\alpha_1}\otimes \ldots\otimes \ket{\alpha_N}\in \mathcal{K}_{N}\,,
\label{eq:product_state}
\ee
where $\mathcal{K}_{N}=h_1\otimes \ldots\otimes h_N$ and $h_j\simeq  \mathbb{C}^8$ is the space generated by $\{\ket{1}\,,\ket{ab}\,,\ldots \,, \ket{abcd}\}$. In this picture, the differential equation~\eqref{eq:diff_equation} is equivalent to a quench problem in $\mathcal{K}_{N}$: the system is prepared in the initial product state
\be
\ket{\overline{\mathcal{U}}(0)}\rangle=\ket{\mathbb{1}}\rangle=\ket{1}\otimes \ket{1}\otimes \ldots \otimes \ket{1}\,,
\label{eq:initial_state}
\ee
and left to evolve according to the differential equation
\begin{align}
\frac{{\rm d}}{{\rm d}t}\ket{\overline{\mathcal{U}}(t)}\rangle= H\ket{\overline{\mathcal{U}}(t)}\rangle\,.
\label{eq:diff_equation_state}
\end{align}
Here, $H$ [not be confused with $H_{\rm SYK}$ in \eqref{eq:q_hamiltonian}] is an operator acting on $\mathcal{K}_N$ which plays the role of the Hamiltonian driving the imaginary-time dynamics. The precise form $H$ in terms of local operators can be derived by computing the action on the basis operators~\eqref{eq:operator_product} of the left multiplication by $L$ in \eqref{eq:replica_hamiltonian}; however, even without doing this explicitly, it is easy to show that $H$ is invariant under any permutation of the sites in $\mathcal{K}_N$. This comes from the fact that the operator $L$ in \eqref{eq:replica_hamiltonian} is left unchanged under the exchange of the pairs $\psi^{\alpha}_{i}\psi^{\beta}_{i}$ and $\psi^{\alpha}_{j}\psi^{\beta}_{j}$ for any choice of $i$ and $j$. Since the initial state \eqref{eq:initial_state} also enjoys the same symmetry, we can conclude that the dynamics of $\ket{\overline{\mathcal{U}}(t)}\rangle$ takes place in the subspace $\mathcal{S}_N\subset \mathcal{K}_N$ which is invariant under arbitrary permutations of the sites. This is of course a great simplification for our problem. The permutational symmetry of the Hamiltonian $H$ is ``emergent'' in the sense that it manifests itself only after taking averages over the Brownian disorder, while the Hamiltonian $H_{\rm SYK}$ in Eq.~\eqref{eq:q_hamiltonian} does not display this symmetry for individual random realizations.

In order to study the dynamics in this subspace, we introduce the basis vectors $\ket{\mathcal{O}_{\vec{n}}}\rangle$ for $\mathcal{S}_N$
\begin{widetext}
\begin{align}
\ket{\mathcal{O}_{\vec{n}}}\rangle =& \ket{n_{1}\,,n_{ab}\,,n_{ac}\,,n_{ad}\,,n_{bc}\,,n_{bd}\,,n_{cd}\,,n_{abcd}}=\frac{1}{\sqrt{N!n_1!n_{ab}!n_{ac}!n_{ad}!n_{bc}!n_{bd}!n_{cd}!n_{abcd}!}} \nonumber\\
\times& \sum_{\pi\in S_N}\pi \underbrace{\ket{1}\otimes \ldots \otimes \ket{1}}_{n_1}\otimes 
\underbrace{\ket{ab}\otimes \ldots \otimes \ket{ab}}_{n_{ab}}\otimes 
\underbrace{\ket{ac}\otimes \ldots \otimes \ket{ac}}_{n_{ac}}\otimes
\ldots \otimes
\underbrace{\ket{abcd}\otimes \ldots \otimes \ket{abcd}}_{n_{abcd}} 
 \pi^{-1}\,,
 \label{eq:basis_perm_inv}
\end{align}
where we used the same notations as in Eqs.~\eqref{eq:mapping}, \eqref{eq:product_state}. Here $\pi$ is the unitary operator associated with a generic element in the symmetric group $S_N$, whose action permutes different sites in $\mathcal{K}_N$. Note that, since the sum runs over all the permutations, not all the elements in the sum are linearly independent. 

The basis vectors~\eqref{eq:basis_perm_inv} of the permutation symmetric space $\mathcal{S}_N$ are labeled by sets of $8$ integers $\{n_j\}$, satisfying $\sum_k n_k=N$, where each integer $n_k$ [with $k=1\,,ab\,, \ldots \,, abcd]$ ``counts'' the number of qudits in the level associated with $k$. In fact, it is possible to employ a more convenient representation, by viewing the state \eqref{eq:basis_perm_inv} as an $8$-mode Fock state generated by bosonic creation operators acting on a vacuum $\ket{\Omega}$. In particular, we have the identification
\begin{align}
\ket{n_1,\ldots,n_{abcd}}= \frac{1}{\sqrt{n_1!\cdots n_{abcd}!}}(a_{1}^\dag)^{n_1}(a_{ab}^\dag)^{n_{ab}}
 \cdots (a_{abcd}^\dag)^{n_{abcd}}\ket{\Omega}\,.
\label{eq:bosonic_representation}
\end{align}
Here, each operator $a_{k}^{\dagger}$ creates a collective mode corresponding to the level associated with $k$. In this language, the initial state~\eqref{eq:initial_state} is written as
$\ket{\overline{\mathcal{U}}(0)}\rangle=\frac{1}{\sqrt{N!}}\left(a_1^{\dagger}\right)^N\ket{\Omega}$.

This representation is particularly convenient, due to the fact that the Hamiltonian $H$ in Eq.~\eqref{eq:diff_equation_state} can be written in terms of the same bosonic operators appearing in Eq.~\eqref{eq:bosonic_representation}:
\be
H = \frac{1}{N^3} \left( -2\binom{N}{4} + \frac{1}{4!}\sum_{r=1}^6 (-1)^{\gamma_r} \left[ X_r^4 - X_r^2 (-6N+8) + 3N(N-2)\right]\right)\,,
\label{eq:otoc hamiltonian_first}
\ee
\end{widetext}
where $X_r$ is a bilinear operator of bosons. The explicit form of $X_r$ is derived in Sec.~\ref{sec:derivation_hamiltonian}, cf. Eqs.~\eqref{eq:X_ab}-\eqref{eq:X_cd}.
A formal solution to the problem of computing $\overline{\mathcal{U}}(t)$ is then obtained as
\be
\ket{\overline{\mathcal{U}}(t)}\rangle=e^{Ht}\ket{\overline{\mathcal{U}}(0)}\rangle=e^{Ht}\frac{1}{\sqrt{N!}}\left(a_1^{\dagger}\right)^N\ket{\Omega}\,.
\label{eq:formal_solution}
\ee 
From its explicit form, one can see that the Hamiltonian $H$ commutes with the operator $\sum_{j=1}^8a^\dag_ja_j$, which ``counts'' the total number of bosonic modes; accordingly, the evolved state \eqref{eq:formal_solution} always belongs to the finite-dimensional Hilbert space generated by the basis vectors  \eqref{eq:bosonic_representation}. However, the dimension of the latter is $D=\binom{N+7}{7}$, which grows as $N^7$, strongly limiting any numerical computation based on a brute force implementation of Eq.~\eqref{eq:formal_solution}. Luckily, it is possible to show that the Hamiltonian $H$ has additional symmetries, which are unveiled by means of an appropriate Bogoliubov transformation 
\be
a_{n}=\frac{1}{\sqrt{8}}C_{n,m}b_m\,,
\label{eq:b_modes}
\ee
further reducing the dimension of the effective Hilbert space explored by the dynamics. The matrix element $C_{n,m}$ are reported in Sec.~\ref{sec:derivation_hamiltonian} [cf. Eq.~\eqref{eq:matrix_transformation}]. Then, from the form of the Hamiltonian $H$ in terms of the modes $b_j$ [cf. Eqs.~\eqref{eq:X_ab b-mode}--\eqref{eq:X_cd b-mode}], one obtains that the number operators
\bea
n_{1,2}:=b_1^\dagger b_1 + b_2^\dagger b_2\,,\ n_{3,4}:=b_3^\dagger b_3 + b_4^\dagger b_4\label{eq:n_12}\\
n_{5,6}:=b_5^\dagger b_5 + b_6^\dagger b_6\,,\ n_{7,8}:=b_7^\dagger b_7 + b_8^\dagger b_8\label{eq:n_56}
\eea
are conserved, namely they commute with $H$. Of course, the initial state can also be expressed in terms of the modes introduced in Eq.~\eqref{eq:b_modes}. Using the explicit form of $C_{n,m}$, we obtain
\begin{align}
\ket{\overline{\mathcal{U}}(0)}\rangle= \frac{1}{\sqrt{N!}\sqrt{8}^N}(b_1^\dag-b_2^\dag-b_3^\dag-b_4^\dag\nonumber\\
+b_5^\dag + b_6^\dag + b_7^\dag - b_8^\dag)^N\ket{\Omega}\,
\label{eq:initial_state_bosonic}
\end{align}
and find that the total conserved number $n_{1,2} + n_{3,4} + n_{5,6}+ n_{7,8}$ is $N$.

As we will see in the next section, these formulas allow us to work with effective Hilbert spaces whose dimensions grow either linearly or quadratically with $N$, and hence to provide numerically exact results for very large system sizes.

\subsection{The OTOC and the tripartite information }
\label{sec:OTOC_INFO}

We now discuss the last step of our method, namely the computation of the matrix elements of the form \eqref{eq:intermediate_1}. Let us consider  the most general OTOC
\bea
\mathcal{F}_{(p,n,m)}(t)=\frac{1}{2^{N/2}} {\rm tr} \left\{\Phi^{(p,n)}(t)\Phi^{(p,m)}(0)\right.\nonumber\\
\left.\Phi^{(p,n)}(t)\Phi^{(p,m)}(0)\right\}\,,
\label{eq:final otoc}
\eea
where we introduced
\begin{align}
\Phi^{(p,n)} &= \psi_{i_1}\cdots\psi_{i_p}\, \psi_{j_1}\cdots\psi_{j_n}\,, \label{eq:otoc operators_1}\\
\Phi^{(p,m)}&= \psi_{i_1}\cdots\psi_{i_p}\, \psi_{k_1}\cdots\psi_{k_m}\,,\label{eq:otoc operators_2}
\end{align}
and where all indices are different, i.e.~the operators have only $p$ Majorana fermions in common. Considering Eq.~\eqref{eq:intermediate_1}, we can expand $\overline{\mathcal{U}}(t)=\overline{ U\otimes U^\ast\otimes U\otimes U^\ast }$ into the basis of operators $\mathcal{O}_{\vec{n}}$ corresponding to the vector \eqref{eq:basis_perm_inv} in ${\rm End}(\mathcal{H}_N^{\otimes 4})$. We obtain
\begin{align}
\mathcal{F}_{(p,n,m)}(t)= &\sum_{\vec{n}}c_{\vec{n}}(t)\bra{L} \left( \mathcal{O}_{\vec{n}} \right) \ket{R} \,,
\label{eq:intermediate_2}
\end{align}
where the sum runs over all the sets $\vec{n}=\{n_{j}\}$ with $j=1,ab,\ldots,abcd$ and $\sum_{j} n_j=N$, while $c_{\vec{n}}(t)$ are the coefficients of $\overline{\mathcal{U}}(t)$ in the basis of the operators $\mathcal{O}_{\vec{n}}$. One can now interpret the sum \eqref{eq:intermediate_2} as the scalar product between an appropriate state $\ket{W_{(p,n,m)}}\rangle\in {\rm End}(\mathcal{H}_N^{\otimes 4})$ and $\ket{\overline{\mathcal{U}}(t)}\rangle$; namely we can write
\be
\mathcal{F}_{(p,n,m)}(t)=\langle\braket{W_{(p,n,m)}|\overline{\mathcal{U}}(t)}\rangle\,.
\label{eq:overlap}
\ee
The whole problem of extracting the numerical value of the OTOC from the knowledge of $\overline{\mathcal{U}}(t)$ then boils down to writing down explicitly $\ket{W_{(p,n,m)}}\rangle$. After this is done, one can straightforwardly compute the overlap \eqref{eq:overlap}. 
 
The derivation of the explicit form of $\ket{W_{(p,n,m)}}\rangle$ is however rather technical, and for this reason we postpone it to Sec~\ref{sec:derivation otocs}. The final result, instead, is extremely simple, and reads
\begin{multline}
|W_{(p,n,m)}\rangle\rangle = \frac{\sqrt{8}^N}{\sqrt{N!}}(-1)^{m(m-1)/2+n(n-1)/2 + nm} \\
\times  (-b_3^\dag)^p(-b_2^\dag)^n(-b_4^\dag)^m(b_1^\dag)^{N-p-n-m} |\Omega\rangle\,,
\label{eq:vec_otoc}
\end{multline}
where $\ket{\Omega}$ and $b_{j}$ were introduced in Eqs.~\eqref{eq:bosonic_representation} and \eqref{eq:b_modes} respectively.

Surprisingly, one can also express the R\'enyi-$2$ entropies entering in the definition of the tripartite information~\eqref{eq:def_tripartite_renyi} in the same form. More precisely, choosing the same conventions as Fig.~\ref{fig:entropy_definition} for the bipartitions of input and output of the evolution operator, one can write
\begin{align}
\exp\left[-S^{(2)}_{AC}(\bar\ell)\right]&=\langle\braket{W_{S^{(2)}_{AC}(\bar\ell)}|\overline{\mathcal{U}}(t)}\rangle\,,\label{eq:w_sac}\\
\exp\left[-S^{(2)}_{AD}(\bar\ell)\right]&=\langle\braket{W_{S^{(2)}_{AD}(\bar\ell)}|\overline{\mathcal{U}}(t)}\rangle\,.\label{eq:w_sad}
\end{align}
Here $\bar\ell$ is the length of $B$ and $D$ (chosen to be of the same size),  while we will use $\ell$ for the length of the regions $A$ and $C$.

Once again, we refer the reader to Sec.~\ref{sec:technical_derivation}, where this is explicitly shown, while in the following we report the final result of this analysis, which gives
\begin{equation}
|W_{S^{(2)}_{AC}(\bar\ell)}\rangle\rangle = \frac{\sqrt{8}^N}{\sqrt{N!}}\frac{1}{2^N}(b_1^\dag-b_2^\dag)^\ell(b_1^\dag-b_4^\dag)^{\bar\ell} |\Omega\rangle\,,
\label{eq:vec_AC}
\end{equation}
and
\begin{equation}
|W_{S^{(2)}_{AD}(\bar\ell)}\rangle\rangle = \frac{\sqrt{8}^N}{\sqrt{N!}}\frac{1}{2^{N/2+\bar\ell}}(b_1^\dag)^\ell(b_1^\dag-b_2^\dag+b_3^\dag-b_4^\dag)^{\bar\ell}|\Omega\rangle\,.
\label{eq:vec_AD}
\end{equation}
Similar formulas could be in principle derived also for more general choices of the bipartitions of input and output. This, however, would introduce additional technical difficulties, so we don't derive them here.

\begin{figure*}
	\includegraphics[width=18cm]{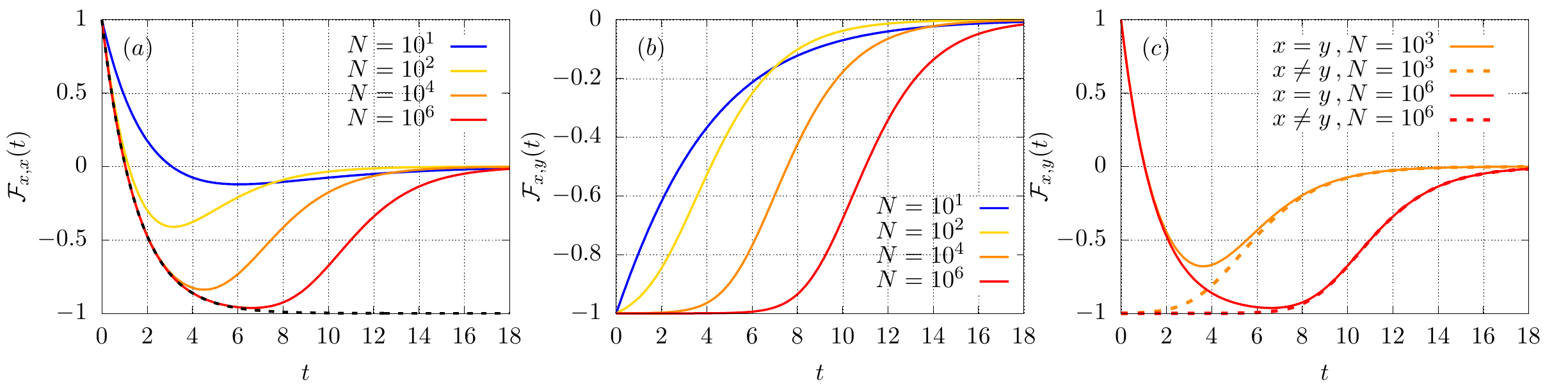}
	\caption{OTOCs $\mathcal{F}_{x,x}(t)$ and $\mathcal{F}_{x,y}(t)$  of single-site Majorana fermions. Subfigures $(a)$ and $(b)$ show our numerical results for increasing system sizes for operators on the same site (from top to bottom) and on different sites (from left to right) respectively. In Subfigure $(a)$ the black dashed line is the analytic prediction \eqref{eq:initial_decay}. Subfigure $(c)$: the two different OTOCs are reported in the same plot, where the  dynamics after the scrambling time is seen to coincide.} 
	\label{fig:OTOC_I}
\end{figure*}

It is now very important to comment on the form of the formulas presented above, as it allows us to reduce the computational cost required to obtain the physical quantities of interest. Let us first consider the case of the OTOC \eqref{eq:final otoc}, which is conveniently rewritten as
\be
\mathcal{F}_{(p,n,m)}(t)=\langle\braket{\overline{\mathcal{U}}(0)|\exp\left(Ht\right)|W_{(p,n,m)}}\rangle^\ast\,.
\label{eq:overlap_evolution}
\ee
Namely, in order to compute $\mathcal{F}_{(p,n,m)}(t)$, one can evolve $|W_{(p,n,m)}\rangle\rangle$ and then take the overlap with the state $\ket{\overline{\mathcal{U}}(0)}$. This is important, as is best appreciated by looking at the simplest instance $p=0,n=m=1$. In this case, Eq.~\eqref{eq:vec_otoc} implies that the state $|W_{(0,1,1)}\rangle \rangle$ belongs to the sector of the Hilbert space labeled by the quantum numbers $n_{1,2} = N-1, n_{3,4} = 1, n_{5,6}=n_{7,8}=0$, where $n_{i,i+1}$ were introduced in Eq.~\eqref{eq:n_12}--\eqref{eq:n_56}. Since $n_{j,j+1}$ are conserved by the Hamiltonian $H$, the dynamics takes place in this sector of the Hilbert space, whose dimension can be easily seen to be $D=N$. Accordingly, one can conveniently represent the restricted Hamiltonian in a basis consisting of $N$ elements, and compute $e^{Ht}|W_{(0,1,1)}\rangle\rangle$ in this basis, which allows us to go to system size one million. 

Similar considerations hold for the generic OTOC $\ket{W_{(p,n,m)}}\rangle$ (which belongs to the sector $n_{1,2}=N-p-m$, $n_{3,4}=p+m$, $n_{5,6}=n_{7,8}=0$) and for the R\'enyi-$2$ entropies corresponding to \eqref{eq:vec_AC}, \eqref{eq:vec_AD}. In the latter cases, expanding
\begin{align}
(b_1^\dag-b_4^\dag)^{\bar\ell}=&\sum_{r=0}^{\bar\ell}\binom{\bar\ell}{r}\left(b_1^\dag\right)^r\left(-b_4^\dag\right)^{\bar\ell-r}\,,\\
(b_1^\dag-b_2^\dag+b_3^\dag & -b_4^\dag)^{\bar\ell}=\sum_{r=0}^{\bar\ell}\binom{\bar\ell}{r}\left(b_1^\dag-b_2^\dag\right)^r\left(b_3^\dag  -b_4^\dag\right)^{\bar\ell-r}\,,
\end{align}
one is left with a sum of terms, each of which requires a simulation within Hilbert spaces up to dimensions $N\bar\ell\sim N^2$. Putting all together, we see that the computation of the quantities of interest requires us to simulate the dynamics in a Hilbert space whose dimension grows either linearly (for the OTOCs) or quadratically (for the tripartite information) with $N$.

\begin{figure*}
	\includegraphics[width=18cm]{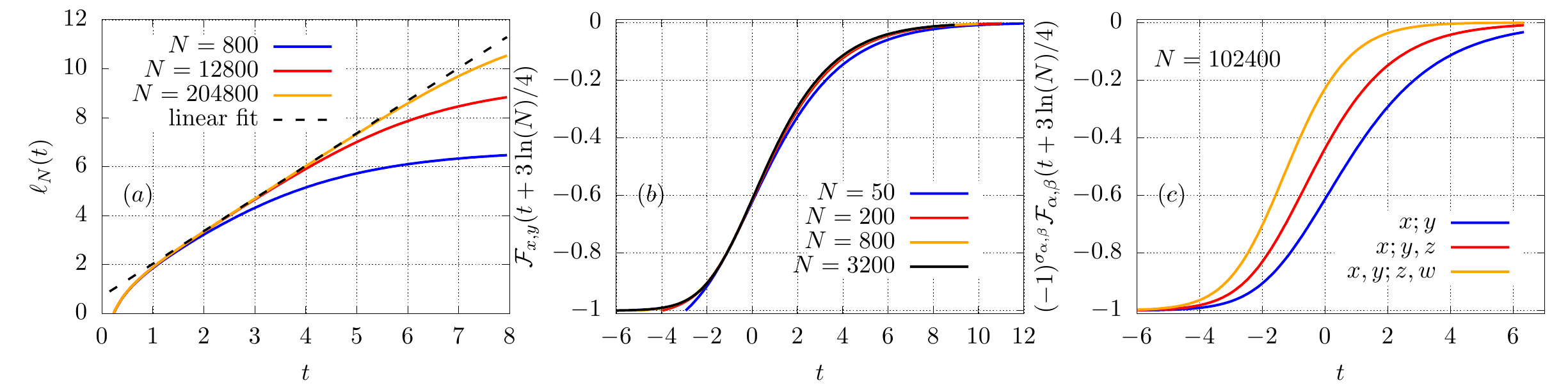}
	\caption{Subfigure $(a)$: rescaled logarithmic OTOC $\ell_N(t)$ [defined in Eq.~\eqref{eq:log_otoc}]. Solid lines correspond to increasing values of $N$ (from bottom to top), while the dashed black line is the linear ansatz $y(t)=\frac{4}{3}t+\ln(2)$. Subfigure $(b)$: data collapse using the shift~\eqref{eq:data_collapse} for the OTOCs $\mathcal{F}_{x,y}(t)$ (with $x\neq y$). Subfigure $(c)$: data collapse for different OTOCs. The curves correspond (from bottom to top) to the OTOCs in Eq.~\eqref{eq:simple_OTOC}, \eqref{eq:OTOC_xyz} and \eqref{eq:OTOC_xyzw} respectively. In order to compare the three curves, we have multiplied $\mathcal{F}_{\alpha,\beta}$ by the global phase $(-1)^{\sigma_{\alpha,\beta}}$, which is $-1$ for $(\alpha;\beta)=(x,y;z,w)$ and $1$ otherwise.}  
	\label{fig:OTOC_II}
\end{figure*}

\section{The physical results}
\label{sec:results}

In this section we present the main physical results of our work.  We begin with the analysis of the OTOCs, and continue with the R\'enyi-$2$ tripartite information introduced in Eq.~\eqref{eq:def_tripartite_renyi}. 

\subsection{The OTOCs: numerical results}

We start by presenting our numerical results for the simplest OTOC 
\be
\mathcal{F}_{x,y}(t) =\frac{1}{2^{N/2}} {\rm tr} \left\{\psi_x(t)\psi_y(0)\psi_x(t)\psi_y(0) \right\}\,.
\label{eq:simple_OTOC}
\ee 
Due to the infinite range of the interactions and the disorder averages, $\mathcal{F}_{x,y}(t)$ does not depend on the precise choice of $x$ and $y$, but only on whether $x=y$ or $x\neq y$. Both cases are displayed in Fig.~\ref{fig:OTOC_I}, where we report data for increasing values of the system size $N$. We see that $\mathcal{F}_{x,x}(t)$ and $\mathcal{F}_{x,y}(t)$ (with $x\neq y$) behave qualitatively differently at short times: the former displays an initial exponential decay, while the latter appears to remain approximately constant. In fact, based on the formulas of the previous section, one can make these statements more precise and show 
\begin{align}
\lim_{N\to\infty}\mathcal{F}_{x,x}(t)=&-1+2\exp\left(-\frac{2}{3}t\right)\,, \label{eq:initial_decay}\\
\lim_{N\to\infty}\mathcal{F}_{x,y}(t)=&-1\,,\label{eq:initial_constant}
\end{align}
where the convergence is point-wise in $t$. This is proven in Sec.~\eqref{sec:derivation limit}. In both OTOCs $\mathcal{F}_{x,x}(t)$ and $\mathcal{F}_{x,y}(t)$, we see the emergence of a characteristic time $t^\ast(N)$, increasing with $N$, which is required before they begin to decay towards zero at large times. One naturally interprets $t^\ast(N)$ as a scrambling time, which is also consistent with our subsequent analysis of the tripartite information. Finally, in Fig.~\ref{fig:OTOC_I}$(c)$ we plot together the OTOCs for $x=y$ and $x\neq y$, for different systems sizes. We see that after an initial time window, the two OTOCs become indistinguishable, meaning that the information on the initial operators chosen has been completely washed out by the chaotic dynamics.

In order to quantitatively characterize the dependence of the scrambling time $t^\ast(N)$ on the system size, we test the short-time behavior of $\mathcal{F}_{x,y}(t)$ against the analytical ansatz~\cite{MaSS16}
\be
\mathcal{F}_{x,y}(t)\sim -1+c_{x,y}\frac{e^{\lambda_{x,y}t}}{N}\,,
\label{eq:ft_ansatz}
\ee
where $c_{x,y}$ is a constant (independent of $N$). In particular, we compute
\be
\ell_N(t)=\ln\left[1+\mathcal{F}_{x,y}(t)\right]+\ln N\,,
\label{eq:log_otoc}
\ee
and compare the numerical data against a linear behavior. The results are shown in Fig.~\ref{fig:OTOC_II}$(a)$. We clearly see that as the system size is increased, the curves for $\ell_N(t)$ approach the linear fit $y(t)=\frac{4}{3}t+\ln(2)$, within an initial time interval that is also increasing with $N$. In turn, this means that the ansatz \eqref{eq:ft_ansatz} is valid, with the free parameters fixed as
\be
\lambda_{x,y}=4/3\,,\qquad c_{x,y}=2\,.
\label{eq:fitted_parameters}
\ee
From this result, we can identify the scrambling time with $t^\ast(N)=3\ln(N)/4$.

\begin{figure*}
	\includegraphics[width=16cm]{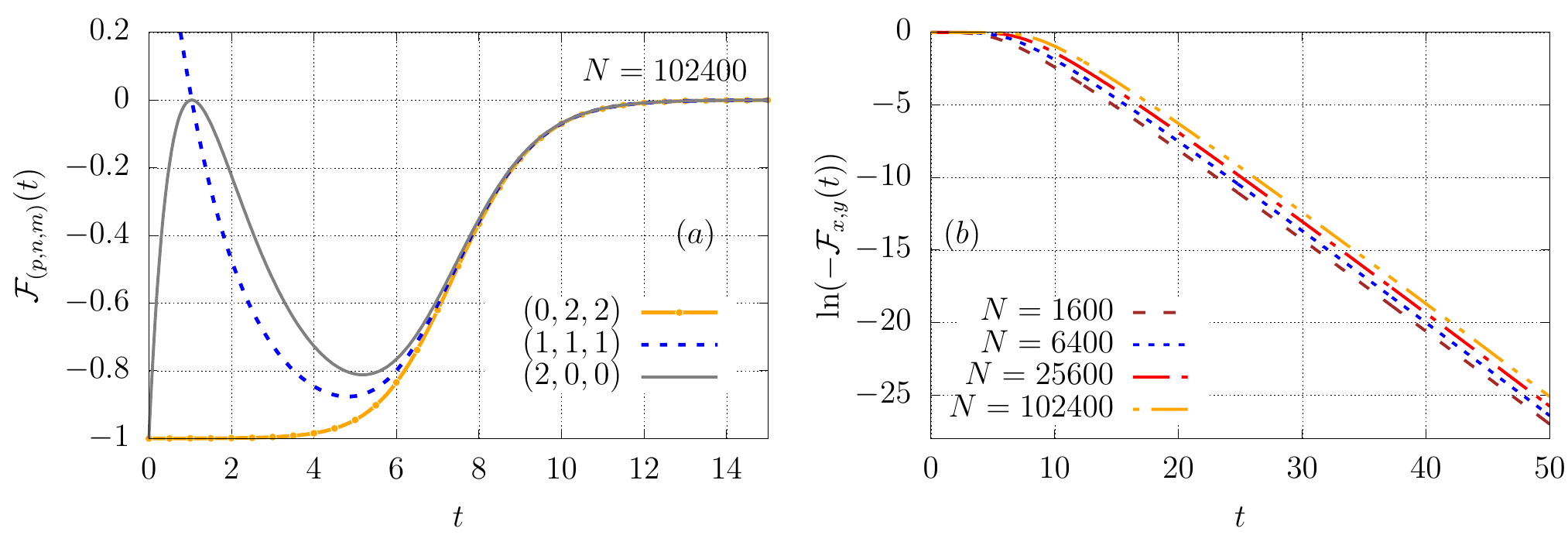}
	\caption{Subfigure $(a)$: time evolution of different OTOCs corresponding to initial strings of Majorana operators with the same length. Each curve is labeled by three integer numbers, according to the convention of Eq.~\eqref{eq:final otoc}. We see that in this case all the OTOCs quickly approach the same curve. Subfigure $(b)$: large-time behavior of the logarithm of the OTOC $\mathcal{F}_{x,y}(t)$. } 
	\label{fig:OTOC_III}
\end{figure*}

The initial behavior in Eq.~\eqref{eq:ft_ansatz}, together Fig.~\ref{fig:OTOC_I}$(b)$, suggests that a data collapse should take place if we consider the shifted functions
\be
\mathcal{F}_{x,y}(t+3\ln(N)/4)\,,
\label{eq:data_collapse}
\ee
where we assumed that the parameters \eqref{eq:fitted_parameters} are exact. This is plotted in Fig.~\ref{fig:OTOC_II}$(b)$, where we see a remarkable data collapse at all times. In particular, the data appear to be perfectly collapsed already for $N\gtrsim 800$.

Next, we have tested how robust the above predictions are, against different choices of the local observables. We have considered in particular
\bea
\mathcal{F}_{x;y,z}(t) =\frac{1}{2^{N/2}} {\rm tr} \left\{\psi_x(t) \psi_{y} (0) \psi_{z} (0)\right.\nonumber\\ \left.\psi_x(t) \psi_{y} (0) \psi_{z} (0)\right\}\,, \label{eq:OTOC_xyz}\\
\mathcal{F}_{x,y; z,w}(t)=\frac{1}{2^{N/2}} {\rm tr} \left\{\psi_{x}(t)\psi_y(t) \psi_z(0)\psi_w(0) \right.\nonumber\\
\left. \psi_{x}(t)\psi_y(t)\psi_z(0)\psi_w(0)\right\}\,.
\label{eq:OTOC_xyzw}
\eea
We have verified that at short times the ansatz~\eqref{eq:ft_ansatz} is always valid, and that a data collapse always takes place using the shift in Eq.~\eqref{eq:data_collapse}. Furthermore the  exponent is universal, namely it is independent of the observables chosen (while the prefactor is not). However, the OTOCs corresponding to distinct choices of local operators are quantitatively different, also after the scrambling time $t^\ast(N)$, as it can be appreciated from Fig.~\ref{fig:OTOC_II}$(c)$. This can be interpreted by saying that, at finite times, the system retains some information on the initial observable chosen.

\begin{figure*}
	\includegraphics[width=18cm]{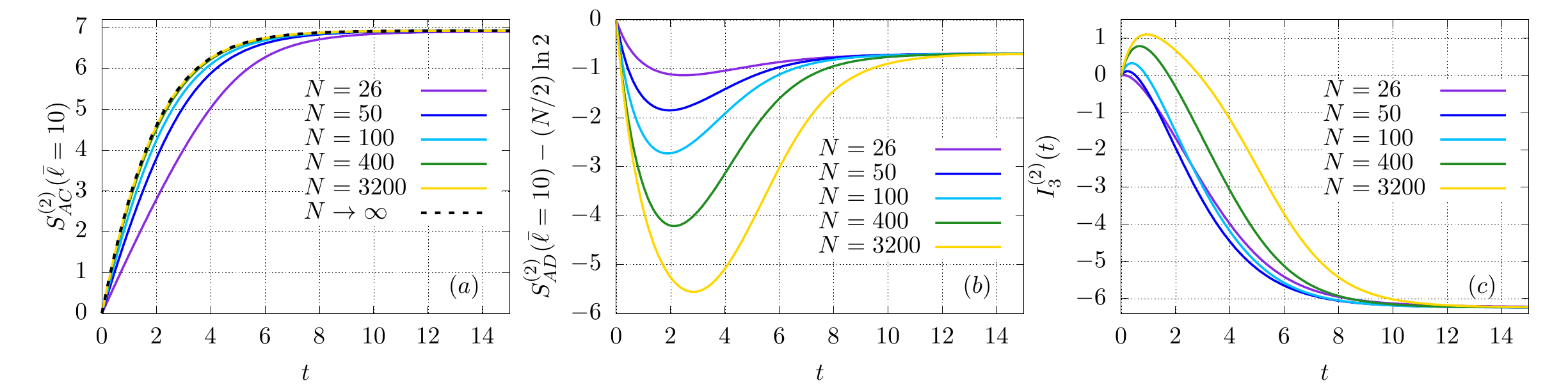}
	\caption{Subfigure $(a)$: time evolution of the R\'enyi-$2$ entropies $S^{(2)}_{AC}$ for the subsystem $A\cup C$, as displayed in Fig.~\ref{fig:entropy_definition}, with $\bar\ell=10$. Solid lines correspond to increasing values of $N$ (from bottom to top), while the black dashed line is the analytic prediction~\eqref{eq:analytic_entropy}. Subfigure $(b)$: time evolution of the R\'enyi-$2$ entropy $S^{(2)}_{AD}$ for the subsystem $A\cup D$, with $\bar\ell=10$. Note that $S^{(2)}_{AD}$ is shifted by its maximum value $(N/2)\ln 2$. Solid lines correspond to increasing values of $N$ (from top to bottom). Subfigure $(c)$: time evolution of the R\'enyi-$2$ tripartite information $I^{(2)}_{3}(A:C:D)$, for the bipartitions $A\cup B$, $C\cup D$ displayed in Figs.~\ref{fig:entropy_definition} and \ref{fig:entropy_majorana} , with $\bar\ell=10$.  Solid lines correspond to increasing values of $N$ (from bottom to top). }  
	\label{fig:tripartite_I}
\end{figure*}

In order to investigate this point further, we plot in Fig.~\eqref{fig:OTOC_III}$(a)$ different OTOCs, corresponding to distinct choices of local observables, which are labeled according to the convention of in Eq.~\eqref{eq:final otoc}. The curves correspond to initial operators that all have the same length, namely that are product of the same number of fermions. In this case, we see that all the OTOCs converge to the same function (up to small corrections) after the scrambling time. Comparing with the results displayed in Fig.~\ref{fig:OTOC_II}$(c)$, we can conclude the following: after the scrambling time, information regarding the specific initial observables is lost, whereas OTOCs corresponding to operators with different initial length can still be distinguished.

Finally, we have investigated the large-time exponential decay of the OTOCs. The data in Fig.~\ref{fig:OTOC_I} suggest to consider an ansatz of the form 
\be
\mathcal{F}_{x,y}(t)\sim d_{x,y}\exp\left[-t/\tau_N\right]\,,
\label{eq:exp_decay}
\ee
where $\tau_N$ should be asymptotically independent of $N$. In Fig.~\ref{fig:OTOC_III}$(b)$, we plot $\ln(-\mathcal{F}_{x,y}(t))$ for large values of $t$, and we see that the data are indeed consistent with an exponential decay of $\mathcal{F}_{x,y}(t)$. To be quantitative, we have performed a fit of $\ln(-\mathcal{F}_{x,y}(t))$ using $r_N(t)=a-t/\tau_{N}-b/t$. For the values of time $t$ available, we have found that the fitted $\tau_{N}$ has a weak dependence on $N$, with $\tau_N\simeq1.53\pm 0.04$ for $N\simeq 10^5$. The fitted value appears to be independent of the choice of the local observables, up to the inaccuracy of the extrapolation method.

\subsection{The R\'enyi-$2$ tripartite information}

We finally present our results for the R\'enyi-$2$ tripartite information introduced in Eq.~\eqref{eq:def_tripartite_renyi}.  As we discussed in Sec.~\ref{sec:OTOC_INFO}, for this quantity the effective dynamics to be computed takes place in a Hilbert space whose dimension grows quadratically with $N$, so that we are restricted to smaller system sizes than in the case of OTOCs. Furthermore, for large subsystems the value of the entropy becomes very large, so that we also have to deal with issues of numerical precision. Overall, for the computationally worst case of bipartitions of equal size, we are able to provide data up to $N\simeq 400$. More details on the numerical implementations are reported in Appendix~\ref{sec:numerical_details}.

\begin{figure*}
	\includegraphics[width=18cm]{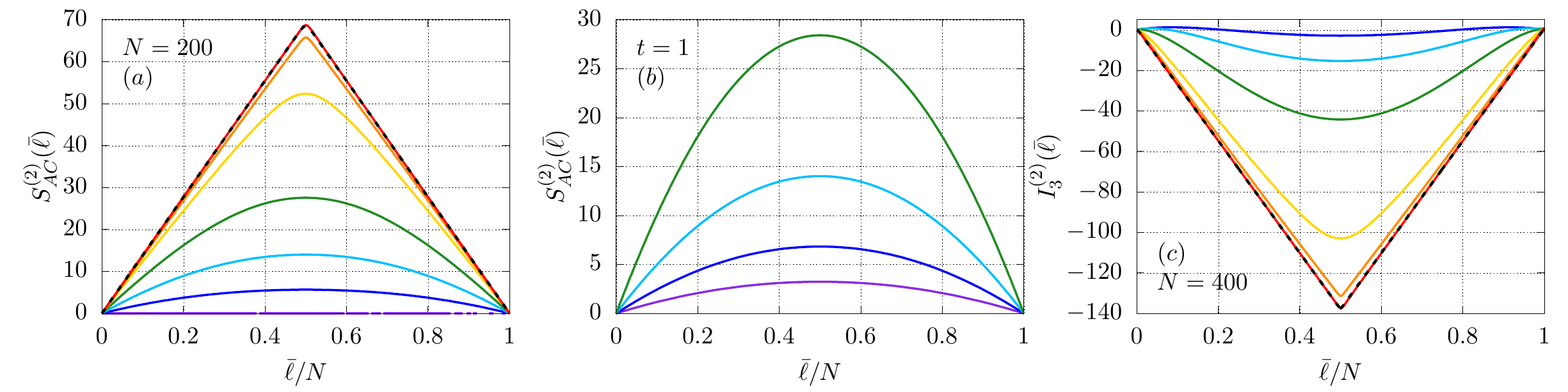}
	\caption{Subfigure $(a)$: R\'enyi-$2$ entropies $S^{(2)}_{AC}$ for the subsystem $A\cup C$, as displayed in Fig.~\ref{fig:entropy_definition}, for $N=200$ and different subsystem sizes $\bar{\ell}$. Solid lines correspond to the times $t=0,0.4,1,2,4,6,15$ (from bottom to top). The black dashed line is obtained by Haar average as computed in Ref.~\cite{HQRY16}. Subfigure $(b)$:  R\'enyi-$2$ entropies $S^{(2)}_{AC}$ computed at $t=1$, for different subsystem sizes $\bar{\ell}$. Solid lines correspond to system sizes $N=50,100,200,400$ (from bottom to top). Subfigure $(c)$: R\'enyi-$2$ tripartite information \eqref{eq:def_tripartite_renyi}  for $N=400$ and different subsystem sizes $\bar{\ell}$. Solid lines correspond to the times $t=0,0.4,1,2,4,6,15$ (from top to bottom). The black dashed line is obtained by Haar average as computed in Ref.~\cite{HQRY16}}  
	\label{fig:tripartite_II}
\end{figure*}

In Fig.~\ref{fig:tripartite_I} we present data for the time evolution of the R\'enyi entropies of the subsystems $A\cup C$ and $A\cup D$, where we used the same partitions as Fig.~\ref{fig:entropy_definition}. The plots correspond to fixed subsystem size and increasing $N$.  Based on the formulas of Sec.~\ref{sec:methods}, in this limit we are able to compute (cf. Sec.~\ref{sec:derivation limit})
\be
\lim _{N \rightarrow \infty, \overline{\ell}, t \text { fix }} S^{(2)}_{A C}(\bar{\ell},t)=\overline{\ell} \ln \frac{2}{1+e^{-2 t / 3}}\,.
\label{eq:analytic_entropy}
\ee
We see from Fig.~\ref{fig:tripartite_I}$(a)$ that the numerical results are in perfect agreement with this prediction. For finite $N$, the entropy $S^{(2)}_{AC}(t)$ displays an initial linear increase, eventually reaching a saturation value, as expected from the traditional picture of quantum quenches. The behavior of the R\'enyi entropy $S^{(2)}_{AD}(t)$ is instead not monotonic, as displayed in Fig.~\ref{fig:tripartite_I}$(b)$. Indeed, one has $S^{(2)}_{AD}(0)=(N/2)\ln 2$, which is the maximum entropy possible, so that at small times $S^{(2)}_{AD}(t)$ has to decrease. Its dynamics is then non-trivial during the initial scrambling time $t^{\ast}(N)$, after which it begins an exponential decay towards its large-time stationary value.

\begin{figure*}
 	\includegraphics[width=16cm]{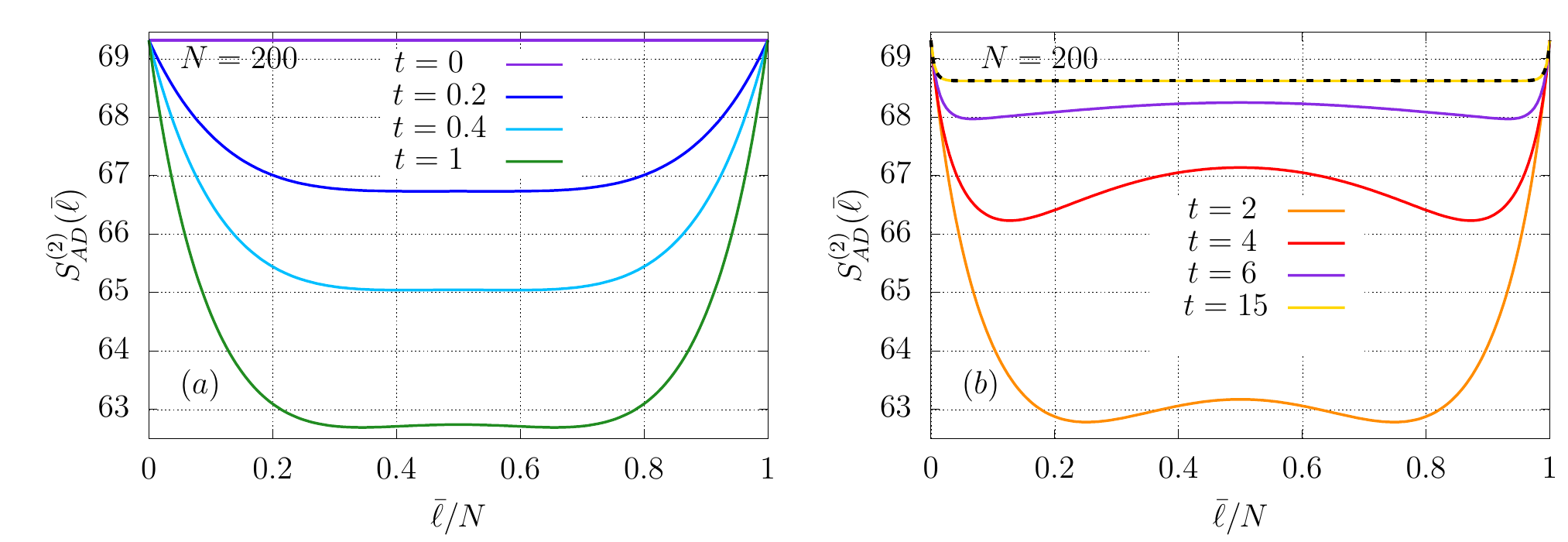}
 	\caption{Time evolution of the R\'enyi-$2$ entropies $S^{(2)}_{AD}$ for the subsystem $A\cup D$, as displayed in Fig.~\ref{fig:entropy_definition}, for $N=200$ and different values of $\bar\ell$. Solid lines in subfigures $(a)$ and $(b)$ correspond, respectively, to times $t=0,0.2,0.4,1$ (from top to bottom) and $t=2,4,6,15$ (from bottom to top). The black dashed line in Subfigure $(b)$ is obtained by Haar average as computed in Ref.~\cite{HQRY16}.} 
 	\label{fig:crossed_entropy}
 \end{figure*}

Figs.~\ref{fig:tripartite_II} and \ref{fig:crossed_entropy} show the same quantities for all the possible values of the subsystems $\bar{\ell}$, at different times and system sizes. First, we notice that the entropies and the tripartite information are symmetric under exchange $\bar{\ell}\leftrightarrow\ell=N-\bar\ell$, as they should. Furthermore, we see that for different values of $\bar \ell$ we have the same qualitative behavior, where at large times an asymptotic value is always reached. In fact, it is possible to compute the latter exactly, as it is known that unitary evolutions driven by Brownian Hamiltonians converge in the infinite-time limit to unitary $k$-designs, for arbitrary positive integers $k$~\cite{BaBK17,OBKB17}. As a consequence, the asymptotic properties can be computed using Haar averages. The latter, which were already computed in Ref.~\cite{HQRY16}, are reported as dashed lines in Fig.~\ref{fig:tripartite_II} and \ref{fig:crossed_entropy}, towards which convergence is apparent. We note that, while their infinite-time limit could be expected, the entropies undergo non-trivial dynamics at short and intermediate times. This is best appreciated by looking at the entropy $S^{(2)}_{AD}(\bar \ell)$ in Fig.~\ref{fig:crossed_entropy}. We see that up to the scrambling time $t^\ast(N)$ it appears to be decreasing (precisely, its average over $\bar{\ell}$), while at later times it increases again. This results in the non-trivial dynamics of the tripartite information, which can become positive at short times [cf. Fig.~\ref{fig:tripartite_I}$(c)$].

\section{Deriving the key formulas}
\label{sec:technical_derivation}

In this last section, we finally address the most technical aspects of our calculations, including several details of the method outlined in Sec.~\ref{sec:methods}. We start by presenting the explicit form of the Hamiltonian driving the dynamics in the four-replica space in Sec.~\ref{sec:derivation_hamiltonian}. Next, we derive the key formulas \eqref{eq:vec_otoc}, \eqref{eq:vec_AC} and \eqref{eq:vec_AD} in Sec.~\ref{sec:derivation otocs}. Finally, in Sec.~\ref{sec:derivation limit} we report the proof of Eqs.~\eqref{eq:initial_decay} and \eqref{eq:analytic_entropy} for the large-$N$ limit of the OTOC $\mathcal{F}_{x,y}(t)$, and of the R\'enyi entropy $S^{(2)}_{AC}(\bar{\ell})$.

\subsection{The Hamiltonian}
\label{sec:derivation_hamiltonian}

In this section we show how to derive the explicit form \eqref{eq:otoc hamiltonian_first} of the Hamiltonian driving the imaginary-time evolution in Eq.~\eqref{eq:diff_equation_state}, from Eq.~\eqref{eq:replica_hamiltonian}. We start with the identity
\begin{align}
4!\sum_{\mathclap{1\le j<k<l<m\le N}} x_ix_jx_kx_l &\nonumber\\
= X^4 -& X^2(-6N+8)+3N(N-2)\,,
\label{eq:identity}
\end{align}
with $X=\sum_{i=1}^N x_i$, for commuting operators $x_i$ satisfying $x_i^2 = -1$. This can be easily derived as follows (see e.g Ref.~\cite{SaSS18}). First, define
\be
f_q=q!\sum_{\mathclap{1\le i_1< \ldots < i_q\leq N}} x_{i_1}\ldots  x_{i_q}\,.
\ee
Then, using $x_j^2=-1$, it is straightforward to show
\be
Xf_q=f_{q+1}-q(N+1-q)f_{q-1}\,,
\ee
which immediately yields the desired identity. Eq.~\eqref{eq:identity} allows us to write the Hamiltonian in terms of global sums of pairs of single-site Majorana operators.  

Next, suppose that for a single-site operator $x_i$ we have
\begin{equation}
x_i \mathcal{O}_i^\alpha = c(\alpha) \mathcal{O}_i^{f(\alpha)}\ \forall \alpha\in\{1,ab,\ldots,abcd\}\,,
\end{equation}
where $\mathcal{O}^\alpha_j$ have been defined after Eq.~\eqref{eq:mapping}. Then one can make the following identification
\begin{equation}
X = \sum_{\alpha=1}^{abcd} c(\alpha) a_{f(\alpha)}^\dag a_\alpha\,,
\label{eq:identification}
\end{equation}
namely the action of $X$ on the permutation symmetric basis \eqref{eq:basis_perm_inv} is the same as the r.h.s. of Eq.~\eqref{eq:identification}, as can be checked directly. From this, the final form of the Hamiltonian in terms of bosonic modes $a_j$ is readily obtained, and reads 
\begin{widetext}
\be
H = \frac{1}{N^3} \left( -2\binom{N}{4} + \frac{1}{4!}\sum_{r=1}^6 (-1)^{\gamma_r} \left[ X_r^4 - X_r^2 (-6N+8) + 3N(N-2)\right]\right),
\label{eq:otoc hamiltonian}
\ee
where $(-1)^{\gamma_r}$ is given in \eqref{eq:gamma_sign}, while the operators $X_r$ are defined as
\begin{align}
X_{ {ab}} =
a^\dag_{ {ab}} a_{ {1}} 
-a^\dag_{ {1}} a_{ {ab}} 
-a^\dag_{ {bc}} a_{ {ac}} 
-a^\dag_{ {bd}} a_{ {ad}} 
+a^\dag_{ {ac}} a_{ {bc}} 
+a^\dag_{ {ad}} a_{ {bd}} 
+a^\dag_{ {abcd}} a_{ {cd}} 
-a^\dag_{ {cd}} a_{ {abcd}}\,, 
\label{eq:X_ab}
\\
X_{ {ac}} = 
a^\dag_{ {ac}} a_{ {1}} 
+a^\dag_{ {bc}} a_{ {ab}} 
-a^\dag_{ {1}} a_{ {ac}} 
-a^\dag_{ {cd}} a_{ {ad}} 
-a^\dag_{ {ab}} a_{ {bc}} 
-a^\dag_{ {abcd}} a_{ {bd}} 
+a^\dag_{ {ad}} a_{ {cd}} 
+a^\dag_{ {bd}} a_{ {abcd}}\,,
\\
X_{ {ad}} = 
a^\dag_{ {ad}} a_{ {1}} 
+a^\dag_{ {bd}} a_{ {ab}} 
+a^\dag_{ {cd}} a_{ {ac}} 
-a^\dag_{ {1}} a_{ {ad}} 
+a^\dag_{ {abcd}} a_{ {bc}} 
-a^\dag_{ {ab}} a_{ {bd}} 
-a^\dag_{ {ac}} a_{ {cd}} 
-a^\dag_{ {bc}} a_{ {abcd}}\,,
\\
X_{ {bc}}  = 
a^\dag_{ {bc}} a_{ {1}} 
-a^\dag_{ {ac}} a_{ {ab}} 
+a^\dag_{ {ab}} a_{ {ac}} 
+a^\dag_{ {abcd}} a_{ {ad}} 
-a^\dag_{ {1}} a_{ {bc}} 
-a^\dag_{ {cd}} a_{ {bd}} 
+a^\dag_{ {bd}} a_{ {cd}} 
-a^\dag_{ {ad}} a_{ {abcd}}\,,
\\
X_{ {bd}}  = 
a^\dag_{ {bd}} a_{ {1}} 
-a^\dag_{ {ad}} a_{ {ab}} 
-a^\dag_{ {abcd}} a_{ {ac}} 
+a^\dag_{ {ab}} a_{ {ad}} 
+a^\dag_{ {cd}} a_{ {bc}} 
-a^\dag_{ {1}} a_{ {bd}} 
-a^\dag_{ {bc}} a_{ {cd}} 
+a^\dag_{ {ac}} a_{ {abcd}}\,,
\\
X_{ {cd}}  = 
a^\dag_{ {cd}} a_{ {1}} 
+a^\dag_{ {abcd}} a_{ {ab}} 
-a^\dag_{ {ad}} a_{ {ac}} 
+a^\dag_{ {ac}} a_{ {ad}} 
-a^\dag_{ {bd}} a_{ {bc}} 
+a^\dag_{ {bc}} a_{ {bd}} 
-a^\dag_{ {1}} a_{ {cd}} 
-a^\dag_{ {ab}} a_{ {abcd}}\,. 
\label{eq:X_cd}
\end{align}

Inspection of Eq.~\eqref{eq:otoc hamiltonian} reveals that the Hamiltonian displays several conservation laws. It is natural to look for a Bogoliubov transformation of the modes which makes some of the symmetries apparent. In addition, one would also like this transformation to simplify the convoluted $a$-mode expression $\ket{\mathcal{W}_{(p,n,m)}}\rangle$ for the OTOCs \eqref{eq:otoc a-modes}. Motivated by this, we look for a transformation where the first boson $b_1$ is associated with the macroscopically occupied mode in Eq.~ \eqref{eq:otoc a-modes}, and choose the other modes $b_j$ to satisfy canonical commutation relations. While this can be done in different ways, it turns out that a particularly convenient transformation is the one defined by Eq.~\eqref{eq:b_modes}, where $C_{n,m}$ is the element in the line $n$ and in the column $m$ of the matrix
\be
C=
\left(
\begin{array}{cccccccc}
	1 & -1 & -1 & -1 & 1 & 1 & 1 & -1 \\
	i & -i & i & i & i & -i & -i & -i \\
	1 & 1 & -1 & 1 & -1 & 1 & -1 & -1 \\
	i & i & i & -i & -i & -i & i & -i \\
	i & i & i & -i & i & i & -i & i \\
	-1 & -1 & 1 & -1 & -1 & 1 & -1 & -1 \\
	i & -i & i & i & -i & i & i & i \\
	-1 & 1 & 1 & 1 & 1 & 1 & 1 & -1 \\
\end{array}
\right)\,.
\label{eq:matrix_transformation}
\ee
Indeed, after this Bogoliubov transformation the form of the Hamiltonian immediately reveals the presence of additional symmetries which can be directly exploited for our computations.

It is straightforward to rewrite the operators~\eqref{eq:X_ab}--\eqref{eq:X_cd}, and hence the Hamiltonian \eqref{eq:otoc hamiltonian}, in terms of the new modes $b_j$. In particular, we have 
\begin{align}
X_{ab} &= i (b_1^\dag b_2+b_2^\dag b_1 + b_3^\dag b_4+b_4^\dag
b_3 -b_5^\dag b_5+b_6^\dag
b_6+b_7^\dag
b_7-b_8^\dag b_8)\,,
\label{eq:X_ab b-mode}\\
X_{ac} &= -b_1^\dag
b_2+b_2^\dag b_1+b_3^\dag
b_4-b_4^\dag b_3 - b_5^\dag b_6+b_6^\dag
b_5+b_7^\dag
b_8-b_8^\dag b_7\,,
\\
X_{ad} &= -i (b_1^\dag
b_1-b_2^\dag b_2-b_3^\dag b_3+b_4^\dag
b_4-b_5^\dag b_6-b_6^\dag
b_5-b_7^\dag
b_8-b_8^\dag b_7)\,,
\\
X_{bc} &= -i (b_1^\dag
b_1-b_2^\dag b_2-b_3^\dag b_3+b_4^\dag
b_4+b_5^\dag b_6+b_6^\dag
b_5+b_7^\dag
b_8+b_8^\dag b_7)\,,
\\
X_{bd} &= b_1^\dag
b_2-b_2^\dag b_1-b_3^\dag b_4+b_4^\dag
b_3-b_5^\dag b_6+b_6^\dag
b_5+b_7^\dag
b_8-b_8^\dag b_7\,,
\\
X_{cd} &= i (b_1^\dag
b_2+b_2^\dag b_1+b_3^\dag
b_4+b_4^\dag b_3+b_5^\dag b_5-b_6^\dag
b_6-b_7^\dag
b_7+b_8^\dag b_8)\,. 
\label{eq:X_cd b-mode}
\end{align}
\end{widetext}

\subsection{Extracting OTOCs and Entropies}
\label{sec:derivation otocs}

We now wish to show how to derive an explicit expression for the vectors $|W_{(p,n,m)}\rangle\rangle$, $|W_{S^{(2)}_{AC}(\bar\ell)}\rangle\rangle$ and $|W_{S^{(2)}_{AD}(\bar\ell)}\rangle\rangle $ in Eqs.~\eqref{eq:vec_otoc}, \eqref{eq:vec_AC} and \eqref{eq:vec_AD}, respectively. In order to simplify this task, we start by proving the following lemma. 
Let
\begin{align}
	|W_N\rangle\rangle &= \sum_{n_1+\cdots+n_{abcd}=N} \frac{1}{\sqrt{N!n_1!\cdots n_{abcd}!}} \\
	&\ \sum_{\pi\in S_N} \pi  \prod_{x=1}^N \alpha_x(z_x) \pi^{-1}
	|n_1,\ldots,n_{abcd}\rangle, \label{eq: W_N}
\end{align}
where $z_x\in\{1,ab,ac,ad,bc,bd,cd,abcd\}$ is the operator at site $x$ for the permutation $\pi$, c.f.~\eqref{eq:basis_perm_inv} and $\alpha_x(z_x)$ constants. Then
\begin{align}
	\ket{W_N}\rangle &= \frac{1}{\sqrt{N!}} \prod_{x=1}^N  \left( \sum_{z=1}^{abcd} \alpha_x(z)a_z^\dag\right) \ket{\Omega}. \label{eq: W_N result}
\end{align}
The equivalence between Eqs.~\eqref{eq: W_N} and \eqref{eq: W_N result} is best established by directly expanding the product in Eq.~\eqref{eq: W_N result}, and regrouping the different terms.

Next, we introduce some notations to handle our subsequent calculations in a compact way. In particular, let us rewrite the basis operator $\mathcal{O}_{\vec{n}}$ in Eq.~\eqref{eq:basis_perm_inv} as
\begin{align}
	\mathcal{O}_{\vec{n}} &= \frac{1}{\sqrt{N!n_1!\cdots n_{abcd}!}}\nonumber\\
	\times & \sum_{π\in S_N}π\, Ψ_{ab}^{ab}Ψ_{ac}^{ac}Ψ_{ad}^{ad}Ψ_{bc}^{bc}Ψ_{bd}^{bd}Ψ_{cd}^{cd}Ψ_{abcd}^{abcd}\, π^{-1}\,.
\end{align}
Here we introduced the notations $\Psi^{ab}_{ab}=\prod_{p \in I_{a b}} \psi_{p}^{a} \psi_{p}^{b}$, $\Psi^{ac}_{ac}=\prod_{p \in I_{a c}} \psi_{p}^{a} \psi_{p}^{c}$, $\ldots $, $\Psi^{abcd}_{abcd}=\prod_{p \in I_{abcd}} \psi_{p}^{a} \psi_{p}^{b}  \psi_{p}^{c} \psi_{p}^{d}$, where $I_{ab}$, $I_{ac}$, $\ldots $ $I_{abcd}$ are ordered, pairwise disjoint, subsets of $\{1,2,\ldots N\}$, such that $|I_{ab}|=n_{ab}$, $\ldots $ $|I_{abcd}|=n_{abcd}$. In this notation, upper indexes in $\Psi^{\alpha}_\beta$ indicate the type of single-site operators, while lower indices specify which subset of $\{1,2,\ldots , N\}$ the product of such operators runs over. Consistent with this convention, we also introduce
\be
\Psi^{a}_a=\prod_{p \in I_{a}} \psi_{p}^{a}\,,
\ee
where $I_{a}$ is the ordered set defined by
\be
I_{a}=I_{a b} \cup I_{a c} \cup I_{a d} \cup I_{a b c d}\,,
\label{eq:ordered_I_a}
\ee
and whose elements are ordered as they appear in Eq.~\eqref{eq:ordered_I_a} (namely, the first $n_{ab}$ elements of $I_{a}$ are those of $I_{ab}$ with the same order, followed by those of $I_{a c}$, and so on). Analogously, one can define $\Psi^{b}_b$, $\Psi^{c}_c$ and $\Psi^{d}_d$, where 
\bea
I_{b}=I_{a b} \cup I_{b c} \cup I_{b d} \cup I_{a b c d}\,, \label{eq:ordered_I_b}\\
I_{c}=I_{a c} \cup I_{b c} \cup I_{c d} \cup I_{a b c d}\,, \label{eq:ordered_I_c}\\
I_{d}=I_{a d} \cup I_{b d} \cup I_{c d} \cup I_{a b c d}\,,\label{eq:ordered_I_d}
\eea
with the elements of $I^a$, $I^b$ and $I^c$ ordered as they appear in Eqs.~\eqref{eq:ordered_I_b}, \eqref{eq:ordered_I_c} and \eqref{eq:ordered_I_d}. Finally, let us consider two disjoint subsets $A\cup B = \{1\ldots N\}$. Then, we define
\begin{align}
	\Psi^{ab}_{abA}=\prod_{p \in I_{ab}\cap A} \psi_{p}^{a}\psi_{p}^{b}\,,\qquad \Psi^{ab}_{abB}=\prod_{p \in I_{ab}\cap B} \psi_{p}^{a}\psi_{p}^{b}\,,\\
	\Psi^{a}_{aA}=\prod_{p \in I_{a}\cap A} \psi_{p}^{a}\,,\qquad \Psi^{a}_{aB}=\prod_{p \in I_{a}\cap B} \psi_{p}^{a}\,,
\end{align}
and analogously for the other cases. Using these notations, we can rewrite

\begin{widetext}
	
	\begin{align}
		\mathcal{O}_{\vec{n}} &= \mathcal{N} \sum_{π\in S_N}π\, Ψ_{abA}^{ab}Ψ_{acA}^{ac}Ψ_{adA}^{ad}Ψ_{bcA}^{bc}Ψ_{bdA}^{bd}Ψ_{cdA}^{cd}Ψ_{abcdA}^{abcd}\, Ψ_{abB}^{ab}Ψ_{acB}^{ac}Ψ_{adB}^{ad}Ψ_{bcB}^{bc}Ψ_{bdB}^{bd}Ψ_{cdB}^{cd}Ψ_{abcdB}^{abcd} π^{-1}\nonumber\\
		&= \mathcal{N} \sum_{π\in S_N}π\,  Ψ_{aB}^aΨ_{bB}^bΨ_{cB}^cΨ_{dB}^d\,Ψ_{aA}^aΨ_{bA}^bΨ_{cA}^cΨ_{dA}^d\,(-1)^{\gamma_A+\gamma_B}π^{-1} \nonumber\\
		&= \mathcal{N} \sum_{π\in S_N}π\, Ψ_{aB}^a Ψ_{aA}^aΨ_{bB}^b Ψ_{bA}^bΨ_{cB}^c Ψ_{cA}^cΨ_{dB}^d Ψ_{dA}^d \,(-1)^{γ_A+γ_B+δ}\pi^{-1}\,,
		\label{eq:basis sorting}
	\end{align}
\end{widetext}
where $\mathcal{N}=(N!n_1!\cdots n_{abcd}!)^{-1/2}$ is the normalization. In order to write down the first line, we used that even string of different fermions commute, while sorting the Majorana operators in the second line resulted in the phases $(-1)^{\gamma_A}$, $(-1)^{\gamma_B}$. We will not write $\gamma_A$, $\gamma_B$ explicitly, as they will cancel at the end of the calculations. Conversely, one can easily compute the phase $(-1)^{\delta}$ appearing in the last line of Eq.~\eqref{eq:basis sorting}:
\begin{align}
	(-1)^\delta &= (-1)^{n_{aA}(n_{bB}+n_{cB}+n_{dB}) + n_{bA}(n_{cB}+n_{dB}) + n_{cA}n_{dB}} \nonumber\\
	&= (-1)^{(n_{aB}^2 + n_{bB}^2 + n_{cB}^2 + n_{dB}^2)/2}\,.
\end{align}
Here we have used that $n_a = n_{aB}+n_{aA}$ and $n_{aB}+n_{bB} + n_{cB} + n_{dB}$ are even, which can be seen by writing explicitly $n_{aB} = n_{abB}+n_{acB}+n_{adB}+n_{abcdB}$ etc.

Eq.~\eqref{eq:basis sorting} is the starting point to derive the explicit form of the vectors $|W_{(p,n,m)}\rangle\rangle$, $|W_{S^{(2)}_{AC}(\bar\ell)}\rangle\rangle$ and $|W_{S^{(2)}_{AD}(\bar\ell)}\rangle\rangle $ for OTOCs and R\'enyi entropies respectively. These are treated in the following, in dedicated subsections.

\subsubsection{The OTOCs}
\label{sec:subsec_otoc}

We wish to calculate the OTOC \eqref{eq:final otoc} of the initial operators \eqref{eq:otoc operators_1}, \eqref{eq:otoc operators_2}.
Starting from \eqref{eq:intermediate_1}-\eqref{eq:L,R otocs}, we can insert (\ref{eq:basis sorting}) for the correlated time evolution operator, where simply $A=\{1\ldots N\}, B=\{\}$ such that there is only one $(-1)^γ$ and no $(-1)^δ$ and we omit the labels $A,B$. Since this expression involves products of even numbers of Majorana fermions acting on the same ``replica'' space, we may switch to the operators $\chi_j^{\alpha}$ in Eqs.~\eqref{eq:chi_1}, \eqref{eq:chi_2}, and perform backwards the steps to derive Eq.~\eqref{eq:intermediate_1} from Eq.~\eqref{eq:general_OTOC}. This gives
\begin{align}
	\mathcal{F}_{(p,n,m)} =& \sum_{\vec{n}} \mathcal{N} c_{\vec n}(t) \sum_{\pi\in S_N}	\pi \tr \left[\Phi^{(p,n)} Ψ_a \right.\nonumber\\
	\Phi^{(p,m)} Ψ_b^\dag &\left. \Phi^{(p,n)} Ψ_c \Phi^{(p,m)} Ψ_d^\dag \right] (-1)^\gamma \pi^{-1}/2^{N/2}\,,
\end{align}
where $ \Phi^{(p,n)}$,  $\Phi^{(p,m)}$ are defined in Eqs.~\eqref{eq:otoc operators_1}, \eqref{eq:otoc operators_2}. Here, we simply wrote $\Psi_{a}$, $\Psi_{b}$, $\Psi_{c}$ and $\Psi_{d}$ without superscript, as we only have a single copy of the fermionic space. More explicitly, we have, for instance
\be
\Psi_a=\prod_{p \in I_{a}} \psi_{p}\,,
\ee
where $I_a$ is defined in \eqref{eq:ordered_I_a}.

Next we move the operator pairs $\Phi^{(p,n)}$ and $\Phi^{(p,m)}$ together such that they cancel. Of course, this generates phases through the anti-commutation relations of the fermions; we obtain
\begin{align}
	\mathcal{F}_{(p,n,m)} = \sum_{\vec{n}} \mathcal{N} c_{\vec n}(t) \sum_{\pi\in S_N} \pi \tr   \left[ {Ψ_aΨ_b^\dag Ψ_cΨ_d^\dag}\right] \nonumber\\
	\times (-1)^\gamma (-1)^{n_{a,c}\left(\{i_\alpha\}\right) + n_{a,b}\left(\{j_\alpha\}\right) +  n_{b,c}\left(\{k_\alpha\}\right)} \nonumber\\
	(-1)^{m(m-1)/2 + n(n-1)/2 + mn} \pi^{-1}/ 2^{N/2}\,,
	\label{eq:otoc_calc_intermediate}
\end{align}
where $n_{a,c}\left(\{i_\alpha\}\right)$ is the number of indeces in $\{i_\alpha\}_{\alpha=1}^p$ which also belong to $I_{a}\cup I_c$ [as defined in Eqs.~\eqref{eq:ordered_I_a}, \eqref{eq:ordered_I_c}]. Analogously,  $n_{a,b}\left(\{j_\alpha\}\right)$ and $n_{b,c}\left(\{k_\alpha\}\right)$ are, respectively, the numbers of indexes in $\{j_\alpha\}_{\alpha=1}^n$ and in $\{k_\alpha\}_{\alpha=1}^m$ which also belong to $I_{a}\cup I_b$ and $I_{b}\cup I_c$. Finally, noticing
\begin{align}
	\tr   \left[ {Ψ_aΨ_b^\dag Ψ_cΨ_d^\dag}\right]= &(-1)^{(n_b + n_d)/2} \tr   \left[ {Ψ_aΨ_b Ψ_cΨ_d}\right]\nonumber\\
	=&2^{N/2} (-1)^{(n_b + n_d)/2} (-1)^\gamma,
\end{align}
we see that the factor $(-1)^\gamma$ in \eqref{eq:otoc_calc_intermediate} is exactly canceled. We are left with an equation of the form \eqref{eq: W_N}, setting $\alpha$'s appropriately. Thus we may apply the lemma proved before [cf. Eq.~\eqref{eq: W_N result}], which directly gives 
\begin{multline}
	|W_{(p,n,m)}\rangle\rangle = \frac{1}{\sqrt{N!}}(-1)^{m(m-1)/2 + n(n-1)/2 + nm} \\
	\times (a_1^\dag-i a_{ab}^\dag + a_{ac}^\dag -i a_{ad}^\dag -i a_{bc}^\dag - a_{bd}^\dag -i a_{cd}^\dag - a_{abcd}^\dag)^p\\
	\times (a_1^\dag +i a_{ab}^\dag-a_{ac}^\dag -ia_{ad}^\dag -i a_{bc}^\dag + a_{bd}^\dag +i a_{cd}^\dag - a_{abcd}^\dag)^n\\
	\times (a_1^\dag - i a_{ab}^\dag - a_{ac}^\dag +i a_{ad}^\dag +i a_{bc}^\dag + a_{bd}^\dag -i a_{cd}^\dag - a_{abcd}^\dag)^m\\
	\times (a_1^\dag +i a_{ab}^\dag + a_{ac}^\dag +i a_{ad}^\dag + i a_{bc}^\dag - a_{bd}^\dag +i a_{cd}^\dag - a_{abcd}^\dag)^{N-p-n-m}\ket{\Omega}. \label{eq:otoc a-modes}\\
\end{multline} 

The result \eqref{eq:vec_otoc} is finally obtained after expressing the operators $a_j^\dag$ in terms of the $b$-modes, introduced in Eq.~\ref{eq:b_modes}.

\subsubsection{The R\'enyi-$2$ entanglement entropy $S^{(2)}_{AC}(\lbar)$}
\label{sec:renyi_ac}

Next, we turn to the task of deriving the vector $|W_{S^{(2)}_{AC}(\bar\ell)}\rangle\rangle$ introduced in Eq.~\eqref{eq:w_sac}, corresponding to the exponential of the second R\'enyi entropy $S^{(2)}_{AC}(\lbar)$. 

As we have explained in detail in Sec.~\ref{sec:intro_tripartite}, in order to compute $S_{AC}^{(2)}(\bar\ell)$, we need to consider the evolved state $U^{a}\ket{I^{ab}}$, where $\ket{I^{ab}}$ was introduced in Eq.~\eqref{eq:I_state_fermions}, while the time evolution operator $U^{a}$ acts only on the ``replica'' space $a$ (cf. Fig.~\ref{fig:entropy_majorana}). For our fermionic system, the reduced density matrix of the union of the disjoint sets $A$ and $C$ can then be written as (see e.g. \cite{FaCa10})
\begin{equation}
	ρ_{AC} = \sum_{F^a_A,F^b_C} \frac{1}{2^{\ell}} (F^a_AF^b_C)^\dag \braket{I^{ab}|U^{a\dag}(F_A^aF_C^b)U^a|I^{ab}}\,.
\end{equation}
Here we denoted by $\{F^a_A\}$, and $\{F^b_C\}$ a complete basis of operators in $A$, $C$ respectively; namely $F^a_A$ and $F^b_C$  take value in all the possible strings of Majorana operators supported in $A$ and $C$. Here, as before, we followed the convention that upper indexes indicate the type of single-site operators, while lower indices specify which subset of $\{1,2,\ldots , N\}$ the product of such operators runs over. Through simple manipulations, we have 
\begin{widetext}
	\begin{align}
		\tr\left[\overline{ ρ_{AC}^2}\right] &= \frac{2^\ell}{2^{2\ell}} \sum_{F_A,F_C}\overline{\underbrace{\bra{I^{ab}}U^{a\dag}}_{\bra{I^{ab}}U^{b}_{-}}F_A^aF_C^bU^a\ket{I^{ab}}\underbrace{\bra{I^{cd}}U^{c\dag}}_{\bra{I^{cd}}U^{d}_{-}}\underbrace{F_C^{d\dagger}F_A^{c\dagger}}_{F_A^cF_C^d(-1)^α}U^c\ket{I^{cd}} } \nonumber\\
		&= \frac{1}{2^{\ell}}\sum_{F_A,F_C} (-1)^α\bra{I^{ab}}\otimes \bra{I^{cd}}\ F_A^a\otimes F_A^{c}\ \overline{ U^a_{+}U^{b}_-U^c_+U^{d}_-}\ F_C^b\otimes F_C^{d}\ \ket{I^{ab}}\otimes\ket{I^{cd}}\,,
	\end{align}
	where $U^{\alpha}_\pm(t)$ are defined in Eq.~\eqref{eq:signed_generator}. From this equation we clearly see that, in complete analogy with the case of the OTOCs, we can write also $\tr\left[\overline{ ρ_{AC}^2}\right]$ in the form $\bra{L} \overline{\mathcal{U}}(t)\ket{R}$. As anticipated, this allows us to apply a procedure similar to the one employed for the OTOCs, and derive the vector $|W_{S^{(2)}_{AC}(\bar\ell)}\rangle\rangle$. In particular, we can use the notations introduced in Sec.~\ref{sec:derivation otocs}, and exploit directly Eq.~\eqref{eq:basis sorting}. This yields straightforwardly
	\begin{align}
		\tr\left[\overline{ ρ_{AC}^2}\right] &= \frac{1}{2^\ell\sqrt{N!n_1!\cdots n_{abcd}!}} \sum_{\vec{n}} c_{\vec{n}}(t) \sum_{\pi\in S_N} π (-1)^{\gamma_A+γ_B+δ}\nonumber\\
		&\qquad \underbrace{\sum_{F_A,F_C}(-1)^α\braket{I^{ab}|F_A^a\, Ψ_{aB}^aΨ_{aA}^aΨ_{bB}^bΨ_{bA}^b\,F_C^b|I^{ab}} \braket{I^{cd}|F_A^{c}\, Ψ_{cB}^cΨ_{cA}^cΨ_{dB}^dΨ_{dA}^d\, F_C^{d}|I^{cd}}}_{(*)}\pi^{-1}\,.
	\end{align}
	
	Next, we compute
	\begin{align}
		(*) &= \sum_{F_A,F_C} \braket{I^{ab}|Ψ_{bB}^bΨ_{bA}^b F_A^aΨ_{aB}^aΨ_{aA}^aF_C^b|I^{ab}}\braket{I^{cd}|Ψ_{dB}^dΨ_{dA}^d F_C^{d\dag}F_A^{c\dag} Ψ_{cB}^cΨ_{cA}^c|I^{cd}} \nonumber\\
		&= \sum_{F_A,F_C} \braket{I^{ab}|(Ψ_{bB}^aΨ_{bA}^a)^{\dag} F_A^aΨ_{aB}^aΨ_{aA}^aF_C^b|I^{ab}}\braket{I^{cd}|F_C^{d\dag} (Ψ_{dB}^cΨ_{dA}^c)^{\dag} F_A^{c\dag} Ψ_{cB}^cΨ_{cA}^c|I^{cd}} \nonumber\\
		&= \sum_{F_A,F_C} \braket{I^{ab}| Ψ_{bA}^{a\dag} Ψ_{bB}^{a\dag} F_A^a Ψ_{aB}^aΨ_{aA}^aF_C^{a\dag}|I^{ab}}\braket{I^{cd}|F_C^cΨ_{dA}^{c\dag}Ψ_{dB}^{c\dag} F_A^{c\dag} Ψ_{cB}^c Ψ_{cA}^c |I^{cd}} \label{eq:crossentropy same up to here}\nonumber\\
		&= \sum_{F_A,F_C} (-1)^{n_{aB}\#F_A + n_{dB}\#F_A}\braket{I|Ψ_{bA}^{\dag} \underbrace{Ψ_{bB}^{\dag}Ψ_{aB}}_{2^{-\lbar/2}\tr Ψ_{bB}^{\dag}Ψ_{aB}} F_A Ψ_{aA} F_C^{\dag}|I} \braket{I|F_C Ψ_{dA}^{\dag} F_A^{\dag} \underbrace{ Ψ_{dB}^{\dag}Ψ_{cB} }_{2^{-\lbar/2}\tr Ψ_{dB}^{\dag}Ψ_{cB}} Ψ_{cA}|I}
	\end{align}
	In \eqref{eq:crossentropy same up to here}, all Majorana operators are in the same system, so we leave away the doubled system label. To extract the traces, note that those are the only operators acting on the region $B$ of the system. Thus the Majorana fermions on those sites $B$ already have to cancel in pairs, or the expectation value $\braket{I|\cdot|I}$ will be zero.  Similarly, the only value of $F_C$ with non-zero contribution has
	\begin{equation}
		Ψ_{bA}^{\dag} F_A Ψ_{aA}F_C^{\dag} = \pm 1 \Rightarrow F_C = \pm Ψ_{bA}^{\dag} F_A Ψ_{aA}\,,
	\end{equation}
	which can be inserted into the second expectation value, canceling the $\pm 1$ and giving
	\begin{align}
		(*) &= \frac{1}{2^{\lbar/2}}\tr \left[Ψ_{bB}^\dagΨ_{aB} \right] \frac{1}{2^{\lbar/2}}\tr \left[Ψ_{dB}^\dagΨ_{cB}\right]\, \braket{I|I}\braket{I| Ψ_{bA}^\dag \sum_{F_A} (-1)^{\#(Ψ_{aA}Ψ_{dA}^\dag)\#F_A} F_A Ψ_{aA} Ψ_{dA}^\dag F_A^\dag Ψ_{cA}|I}\nonumber \\
		&= 2^{-\bar\ell}\, \tr \left[Ψ_{bB}^\dagΨ_{aB}\right] \tr\left[Ψ_{dB}^\dag Ψ_{cB}\right] \tr \left[Ψ_{aA}Ψ_{dA}^\dag \right]\tr\left[ Ψ_{bA}^\dag Ψ_{cA}\right] \nonumber\\
		&= 2^{-\bar\ell}(-1)^{\frac{n_b + n_d}{2} + n_{bB}+n_{dB}}  \tr \left[Ψ_{aB}Ψ_{bB}\right] \tr \left[Ψ_{cB}Ψ_{dB}\right] \tr \left[ Ψ_{aA}Ψ_{dA} \right] \tr \left[Ψ_{bA}Ψ_{cA}\right]\,.	\label{eq:end_star}
	\end{align}
\end{widetext}
Here, in order to go from the first to the second line, we made use of the identity \eqref{eq:identity_majorana} in Appendix~\ref{sec:app_majorana_case}. The last line of \eqref{eq:end_star} is non-zero only for
\begin{equation}
	\begin{aligned}
		n_1 = &n_{1B}+n_{1A}\,,\quad  n_{ab} = n_{abB}\,, \quad n_{ac}=0\,,\\
		n_{ad}=& n_{adA}\,,\quad  n_{bc} = n_{bcA}\,, \quad n_{bd} = 0\,, \\
		n_{cd} = & n_{cdB}\,, \quad n_{abcd} = n_{abcdB} + n_{abcdA}\,. 
		\label{eq:entropy ns}
	\end{aligned}
\end{equation}
With this we see that the traces evaluate as $(-1)^{γ_B+γ_A}$. Also, it shows that $n_{aB} \equiv n_{bB} \equiv n_{bA} \equiv n_{cA} \equiv n_{cB} \equiv n_{dB}\ ({\rm mod}\ 2)$, such that $(-1)^\delta = +1$. Putting all together, we get
\begin{align}
	\tr \overline{ρ^2_{AC}} &= \sum_{\vec n}\frac{c_{\vec n}(t)}{\sqrt{N!n_1!\cdots n_{abcd}!}}\nonumber\\
	&\times \sum_{π\in S_N}π\, (-1)^{(n_b+n_d)/2} δ(\pi) π^{-1}
\end{align}
where the Kronecker delta $δ(\pi)$ enforces the constraints~\eqref{eq:entropy ns}. This expression can also be cast into the form \eqref{eq: W_N} by setting some $\alpha$'s to zero. Then Eq.~\eqref{eq: W_N result} gives us
\begin{align}
	|W_{S^{(2)}_{AD}(\bar\ell)}\rangle\rangle= \frac{1}{\sqrt{N!}}\left[ia^{\dagger}_{ab}+ia^{\dagger}_{cd}+(a^{\dagger}_1-a^{\dagger}_{abcd})\right]^{\ell} \nonumber\\
	\left[ia^{\dagger}_{ad}+ia^{\dagger}_{bc}+(a^{\dagger}_1-a^{\dagger}_{abcd})\right]^{\bar \ell}\ket{\Omega}\,.
\end{align}
Transformation to $b$-modes \eqref{eq:b_modes} finally yields the result anticipated in Eq.~\eqref{eq:vec_AC}.

An analogous treatment can be carried out for the case of the entropy $S^{(2)}_{AD}$. Since the technical steps are very similar, we report them in Appendix~\ref{sec:renyi_ad}.

\subsection{Some large-$N$ limits}
\label{sec:derivation limit}

In this section, we finally show how one can compute the limit $N\to\infty$, while keeping time $t$ fixed, for the OTOCs and the R\'enyi-$2$ entropies, and derive in particular Eqs.~\eqref{eq:initial_decay} and \eqref{eq:analytic_entropy}. 

We start with the case of OTOCs, and consider Eq.~\eqref{eq:overlap_evolution}. As a first simplification, we only need to keep modes $b_1$ through $b_4$ in the Hamiltonian and initial state $\langle\bra{\overline{\mathcal{U}}(0)}$ as the others are not present in $\ket{W_{(p,n,m)}}\rangle$. Next, we switch to ladder operators with an unusual normalization, specifically
\begin{equation}
	\tilde b_i^\dag \ket{n_i}_{\tilde b} = \ket{n_i+1}_{\tilde b},\ \tilde{b}_i\ket{n_i}_{\tilde b} = n_i\ket{n_i-1}_{\tilde b}. \label{eq: btilde}
\end{equation}
This now allows us to take the leading order in $N$ for each term of the exponential $e^{Ht}$, using that $\tilde b_1 \sim N$ as $p,n,m\ll N$. We obtain
\begin{align}
	&\lim_{N\to\infty} \langle\braket{\overline{\mathcal{U}}(0)|e^{Ht}|W_{(p,n,m)}}\rangle \\
	&\quad= \sum_{m=0}^\infty \langle\braket{\overline{\mathcal{U}}(0)|\left(\lim_{N\to\infty} H\right)^m|W_{(p,n,m)}}\rangle t^m/m!\,,
	\label{eq:exchanging limits}
\end{align}
with
\begin{gather}
	\lim_{N\to\infty} H = H_A + H_B + H_C, 	\label{eq:asymptotic hamiltonian}\\
	H_A = \frac{2}{3}(\tilde b_2^{\dag 2}\tilde b_1^2/N^2 - 1)\tilde b_2^\dag \tilde b_2,\\ H_B = \frac{2}{3N^3}\tilde b_2^{\dag 3}\tilde b_1^3\tilde b_4^\dag \tilde b_3,\\
	H_C =  - \frac{2}{3}\tilde b_3^\dag \tilde b_3\,.
\end{gather}
As $\langle\bra{\overline{\mathcal{U}}(0)}(\tilde b_2^{\dag 2}\tilde b_1^2/N^2 - 1) = 0$ at the highest order in $N$, terms with $H_A$ do not contribute at the leading order.
For the OTOC $\mathcal{F}_{x,y}(t)$, also $H_B$ and $H_C$ cannot occur, because $\ket{W_{x,y}}\rangle$ \eqref{eq:vec_otoc} does not contain any $b_3$-modes. The asymptotic result is then the constant $\mathcal{F}_{x,y}(t) \to \langle\braket{\overline{\mathcal{U}}(0)|1|W_{x,y}}\rangle = -1$, as reported in \eqref{eq:initial_constant}.
In contrast, for the OTOC $\mathcal{F}_{x,x}(t)$, the state $\ket{W_{x,x}}\rangle$ does contain one $b_3$-mode such that $H_B$ can appear at most once. The remaining Hamiltonian is still simple enough to finally derive the exponential decay \eqref{eq:initial_decay}. We stress that we can only expect these limits to be  point-wise in $t$ due to the exchange of limits in \eqref{eq:exchanging limits}; in fact, convergence is clearly not uniform, as can be seen from the exact numerical results.

The case of the entropy $S_{AC}^{(2)}(t)$ is treated along similar lines. We first perform a further mode transformation
\begin{align}
	c_1 &= (b_1-b_2)/\sqrt{2} \,,& c_2 &= (b_1+b_2)/\sqrt{2}\,, \nonumber\\
	c_3 &= (b_3+b_4)/\sqrt{2}\,, & c_4 &= (b_4-b_3)/\sqrt{2}\,,
	\label{eq:c-modes}
\end{align}
such that
\begin{equation}
	\ket{W_{S^{(2)}_{AC(\bar\ell)}}}\rangle = \frac{2^N}{\sqrt{N!}}\frac{1}{2^{\bar\ell}} (c_1^\dag)^\ell(c_1^\dag+c_2^\dag-c_3^\dag-c_4^\dag)^{\bar\ell}\ket{\Omega}\,.
\end{equation}
We may now follow the same procedure as for the OTOCs. In fact, the Hamiltonian has the exact same form in terms 
of the modes $b_j$ and $c_j$. Taking $\bar\ell\ll N$,  Eq.~\eqref{eq:asymptotic hamiltonian} is therefore valid, after substituting the modes $\tilde b_j$ with $\tilde c_j$. Now, the initial state 
\begin{equation}
	\langle\bra{\overline{\mathcal{U}}(0)} = \bra{\Omega}(c_1-c_3)^N\frac{1}{\sqrt{N!}2^N}
\end{equation}
annihilates both $H_A$ and $H_B$ ensuing in a very simple (quadratic) asymptotic Hamiltonian $H_C$. From this, Eq.~\eqref{eq:analytic_entropy} follows straightforwardly.

\section{Conclusions}
\label{sec:conclusions}

In this work, we have developed an approach to analyze the chaotic dynamics in the Brownian SYK model, a system of $N$ Majorana fermions coupled together via random, time-dependent interactions. We have shown that the OTOCs and the tripartite information of the unitary evolution can be studied as a quench problem (at imaginary times) in a system of $N$ qudits, which can be conveniently investigated in terms of bosonic modes, due to an emergent permutational symmetry. Exploiting the latter, we were able to produce numerically exact results up to $N= 10^6$, and to study several features of the chaotic dynamics at finite size.

We have analyzed in detail the dependence of the OTOCs on the observables chosen, highlighting the pieces of information on the initial operators which are not washed out by the chaotic dynamics. In particular, after the scrambling time $t^\ast(N)\sim \ln N$,  the OTOCs of distinct operators converge to the same curve if they have the same length, namely if they are written as products of the same number of Majorana fermions, whereas the curves of different OTOCs can be distinguished after the scrambling time $t^\ast(N)$ if the length is different. Furthermore, we have verified that the exponent of the initial exponential growth of the OTOCs is universal and performed a data collapse for increasing system sizes. Regarding the tripartite information, we have shown that its evolution is non-trivial during the initial scrambling time, while at large times it always decays exponentially to the corresponding Haar-scrambled value; this result is consistent with the rigorous recent findings of Refs.~\cite{BaBK17,OBKB17}

The approach developed in this paper can be generalized to other models where the Hamiltonian displays all-to-all random interactions, with time-dependent Brownian disorder. Indeed, one can straightforwardly follow the steps outlined in Sec.~\ref{sec:methods}, and study the dynamics of OTOCs and tripartite information as a quench problem in a qudit system with site permutational symmetry. In turn, this implies that the effective imaginary-time dynamics takes place in a Hilbert space whose dimension grows as a polynomial in $N$. Of course, one would need to investigate for each case whether a further reduction of the effective dimension takes place, as for the Brownian SYK model studied in this paper.

It is possible that the final formulas obtained with our method (which have been used in this work mainly for efficient numerical computations) could be simplified further and evaluated to exact analytic expressions in the large-$N$ limit. In fact,  by means of a different approach, an exact result for a suitable average of OTOCs was found in Ref.~\cite{ZhCh19} for the Brownian dynamics generated by a disordered Hamiltonian in a qudit system. It would be interesting to see whether ideas related to the work~\cite{ZhCh19} could be used here, to obtain analytic expressions for the OTOCs of arbitrary observables and for the tripartite information, in the large-$N$ limit. 

Finally, the approach presented in this paper could also be applied to compute quantities involving higher moments of the evolution operator $U(t)$, such as R\'enyi entropies of higher order, or the R\'enyi-$2$ operator entanglement entropy of local observables~\cite{PrPi07,Duba17}. In these cases, however, the application of our method would be inevitably more complicated. More importantly, it is not granted that a reduction of the Hilbert-space dimension could be achieved by means of a transformation analogous to \eqref{eq:b_modes}. In any case, it would be certainly interesting to investigate these points further.

\begin{acknowledgments}

LP thanks Bruno Bertini and Tomaz Prosen for discussions related to this work. XLQ thanks the helpful discussions with David Huse and Alex Streicher. LP acknowledges support from the Alexander von Humboldt foundation. JIC and NS acknowledge support by the EU Horizon 2020 program through the ERC Advanced Grant QENOCOBA (No. 742102, JIC) and the ERC Starting Grant WASCOSYS (No. 636201, NS), and from the DFG (German Research Foundation) under Germany’s Excellence Strategy - EXC-2111 - 390814868. XLQ is supported by the National Science Foundation under grant No. 1720504, and in part by the Department of Energy under grant No. DE-SC0019380.

\end{acknowledgments}

\appendix

	\section{Non-interacting case: $q=2$}
	\label{sec:non-interacting}
	In this section, we study the Brownian SYK model \eqref{eq:q_hamiltonian} for $q=2$. We choose the constant $\sigma_J$ in \eqref{eq:sigma_J} such that the disorder's correlations are given by
	\begin{equation}
	\overline{J_{ij}(t)J_{i'j'}(t')} = \delta_{ii'}\delta_{jj'}\delta(t-t')\frac{1}{N}\,.
	\end{equation}
	Each disorder realization is governed by a free Hamiltonian, therefore we do not expect any scrambling of operators or decay of OTOCs.
	
	The method developed in this article can be applied to arbitrary $q$ and we may study the non-interacting case within its framework. The states 
	\begin{equation}
	\ket{W_{(p,n,m)}}\rangle,\ \ket{W_{S^{(2)}_{AC}(\bar\ell)}}\rangle,\  \ket{W_{S^{(2)}_{AD}(\bar\ell)}}\rangle,\ \text{and}\  \ket{\overline{\mathcal U}(0)\rangle}
	\end{equation}
	representing the OTOC \eqref{eq:vec_otoc}, Rényi-2 entanglement entropies \eqref{eq:w_sac} and \eqref{eq:w_sad}, and initial time evolution operator \eqref{eq:initial_state_bosonic} are independent of $q$ as long as $q$ is even. However, the effective Hamiltonian reflects the change of $q$ and is simpler. Along the same lines as for $q=4$ (see section~\ref{sec:generator_dynamics}), we can compute
	\begin{gather}
		\frac{\mathrm{d}}{\mathrm{d}t} \overline{\mathcal U(t)} = L\overline{\mathcal U(t)},\\
		L = \frac{1}{N}\left[-2\binom{N}{2} - \sum_{\substack{\alpha,\beta =a,b,c,d \\\alpha<\beta}} \sum_{i<j} (ψ_i^\alpha ψ_i^\beta)(ψ_j^\alpha ψ_j^\beta)\right].
	\end{gather}
	
	The corresponding representation of the effective Hamiltonian after operator-state mapping in bosonic modes is
	\begin{align}
		&{\qquad} \ket{\overline{\mathcal U}(t)}\rangle = e^{Ht}\ket{\overline{U}(0)}\rangle, \\
		H &{}= \frac{1}{N}\left[-2\binom{N}{2} -3N - \frac{1}{2}\sum_{r=ab}^{cd} X_r^2\right] \\
		&{}= -\frac{4}{N}(b_1^\dag b_3^\dag - b_2^\dag b_4^\dag)(b_1 b_3-b_2b_4),
	\end{align}
	where the six $X_r$ operators are the same as in the corresponding expression \eqref{eq:otoc hamiltonian_first} for $q=4$.
	
\begin{figure}
		\includegraphics[width=8.5cm]{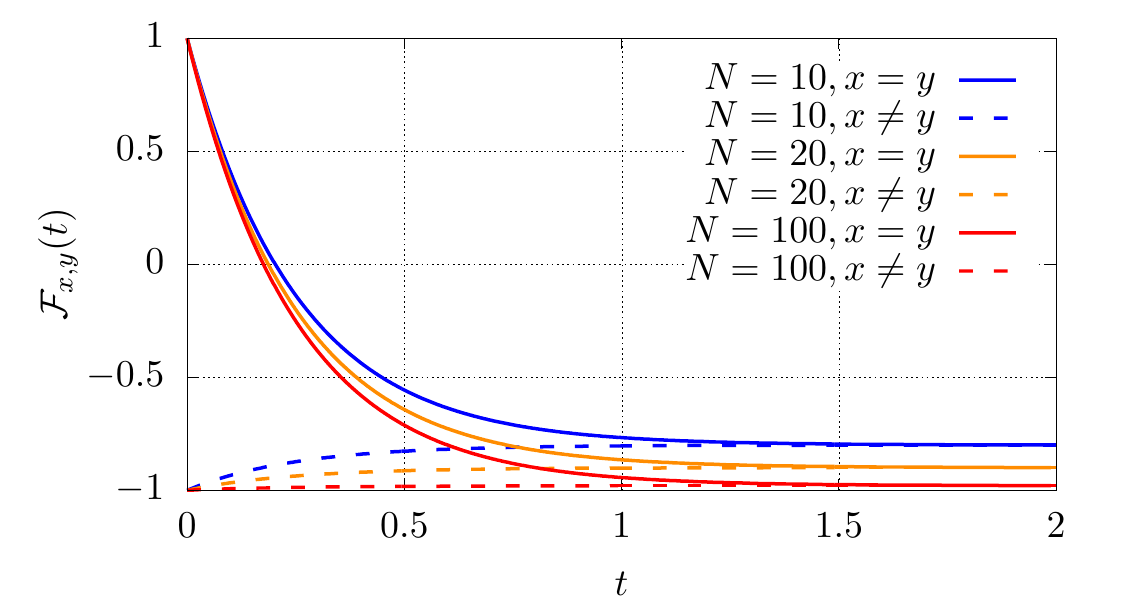}
		\caption{OTOCs $\mathcal{F}_{x,x}(t)$ (solid lines) and $\mathcal{F}_{x,y}(t)$ (dashed lines) for single site Majorana fermions. We show the analytical results \eqref{eq:q=2 xx} and \eqref{eq:q=2 xy} for various system sizes $N$. At long times, they decay to the same value $-1+\frac{2}{N}\neq0$, indicating the absence of scrambling.}
		\label{fig:otocs q=2}
	\end{figure}

	For the two simple OTOCs $\mathcal{F}_{x,y}(t)$ and $\mathcal{F}_{x,x}(t)$ the dynamics only explores the two-level subspace spanned by $\ket{N-1,0,1,0}$ and $\ket{N-2,1,0,1}$. Therefore we can compute these OTOCs analytically and obtain
	\begin{align}
		\mathcal{F}_{x,y}(t) &{}= -1+\frac{2}{N} - \frac{2}{N} e^{-4t}\,, \label{eq:q=2 xy}\\
		\mathcal{F}_{x,x}(t) &  = -1+\frac{2}{N} + \frac{2}{N} e^{-4t} (N-1)\,. \label{eq:q=2 xx}
	\end{align}
	The curves are plotted in Fig.~\ref{fig:otocs q=2}. As expected, the OTOCs do not decay to zero at long times, as the non-interacting model is not chaotic. Since the tripartite information can be written as an average of OTOCs \cite{HQRY16}, it too will lack the characteristics of scrambling.

	We now make a comment on the so-called ``length'' of the operator $ψ_j(t)$, see e.g.~\cite{RoSS18}. At any  time $t$, we can always write
	\begin{equation}
		ψ_j(t) = \sum_s \sum_{\{k_j\}}\underbrace{ψ_{k_1}\cdots ψ_{k_s}}_{\text{length}\ s}c_{s,\{k_j\}}(t)\,,
	\end{equation}
	and define the average length $L(t)$ as
	\begin{equation}
	L(t) = \sum_s s \sum_{\{k_j\}} |c_{s,\{k_j\}}|^2\,.
	\end{equation}
	It can be shown that the average length is related to an appropriate average over OTOCs, namely~\cite{RoSS18}
	\begin{equation}
		L(t) = \frac{N+\mathcal{F}_{x,x}(t) + (N-1)\mathcal{F}_{x,y}(t)}{2}.
	\end{equation}
Using this relation, if follows from our results \eqref{eq:q=2 xy}--\eqref{eq:q=2 xx} that the length is constant 1. This is expected because the Gaussian dynamics preserves the length of products of Majorana operators.
	
\begin{figure*}
		\includegraphics[width=18cm]{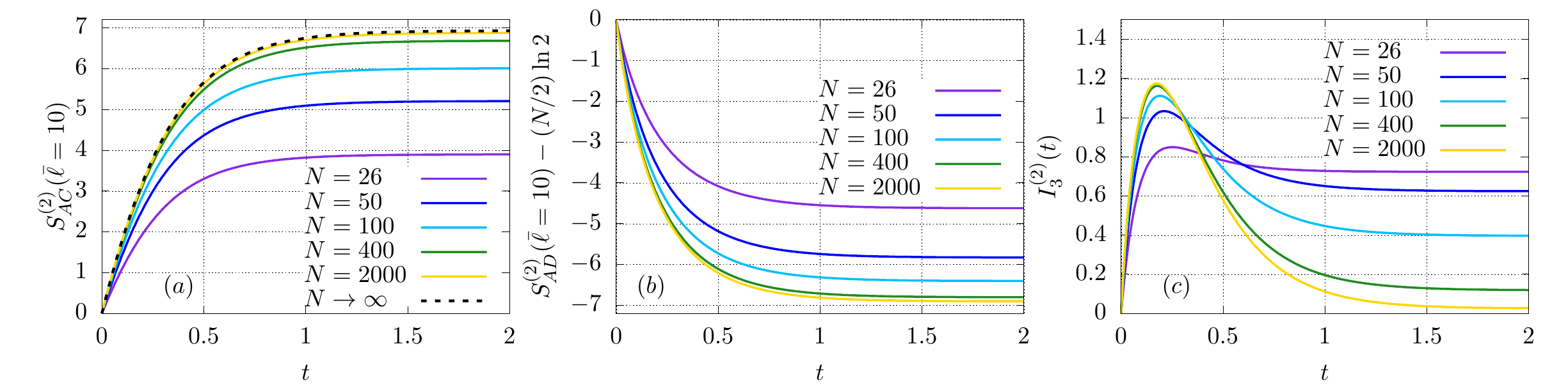}
		\caption{For the free case ($q=2$), we show the time behavior of the entanglement entropies $S_{AC}^{(2)}$ (a), $S_{AD}^{(2)}$ (b) and tripartite information $I_3^{(2)}$ (c) for several system sizes $N$ and fixed subsystem size $\lbar=10$. In (a), we also indicate the limit \eqref{eq:free entropy limit}. The tripartite information is always positive, which means that the non-interacting system does not scramble quantum information.}
		\label{fig:tripartitefree}
	\end{figure*}

	Next, we can calculate the entanglement entropies numerically, just like in the interacting case. We present the results in Fig.~\ref{fig:tripartitefree}. 
	In the limit $N\to\infty, \lbar, t$ fixed, we can derive
		\begin{equation}
		\lim_{N\to\infty, \lbar,t\ \text{fix}} S_{AC}^{(2)}(\lbar,t) = \lbar \ln \frac{2}{1+e^{-4t}},
		\label{eq:free entropy limit}
	\end{equation}	
	along the same lines as in section~\ref{sec:derivation limit}. While the entropy $S_{AC}^{(2)}(\lbar)$ saturates to its maximal Haar value at large $N$ and $t$, the behavior of $S_{AD}^{(2)}(\lbar)$ is qualitatively different from the interacting case (Fig.~\ref{fig:tripartite_I}). This leads to the tripartite information being positive at all times and system sizes, indicating the absence of scrambling.

\begin{figure*}
		\includegraphics[width=14cm]{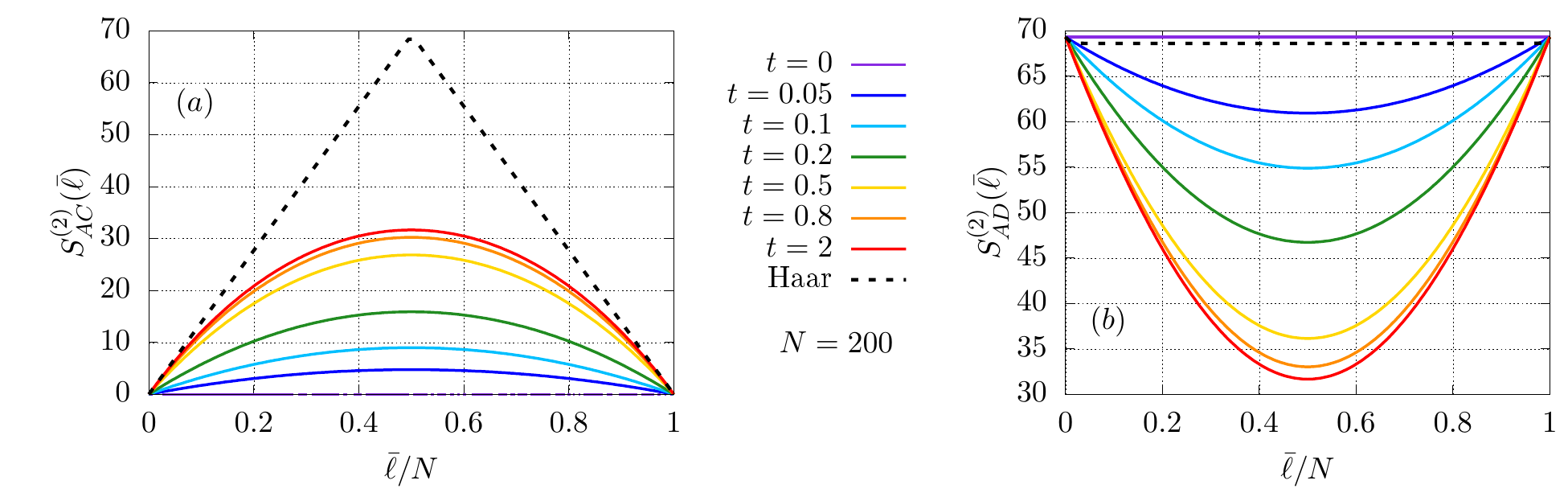}
		\caption{The entanglement entropies $S_{AC}^{(2)}$ (a) and $S_{AD}^{(2)}$ (b) for various subsystem sizes $\lbar$ and several times $t$ in the free case. The black dotted lines show the values reached with Haar-scrambling.}
		\label{fig:freebylbar}
	\end{figure*}

	As for the interacting case, we can also study the entanglement entropies' dependence on the subsystem size, see Fig.~\ref{fig:freebylbar}. Comparing against Figs.~\ref{fig:tripartite_II} and \ref{fig:crossed_entropy}, we see that in the free case, the entanglement entropies do not reach the maximal Haar scrambled values at finite ratios $\bar{\ell}/N$.

\section{Relation between OTOCs and R\'enyi-$2$ entropies}
\label{sec:tripartite_and_otocs}

In this appendix we review the relation between OTOCs and R\'enyi-$2$ entropies in the case of unitary evolution operators defined on qubit systems, and generalize the latter for fermionic (Majorana) systems. We focus on the configuration displayed in Fig.~\ref{fig:entropy_definition}, taking the regions $A$ and $C$ of the same size and position (and analogously for $B$ and $D$). 

\subsection{The case of Pauli matrices}

We start by giving a derivation of the aforementioned relation for a system of qubits, along the lines of the one in Ref.~\cite{HQRY16}. First, we can write the reduced density matrix $\rho_{AC}$ as
\be
\rho_{AC}=\frac{1}{2^{a+c}}\sum_{j,k}\left(O_j^{A}O_k^{C}\right)^{\dagger}\braket{I|U^{\dagger}_{AB}O_j^{A}O_k^{C}U_{AB}|I}\,,
\ee
where the sum is over the complete bases $\{O_j^{A}\}$ and $\{O_k^{C}\}$ of strings of Pauli operators in $A$ and $C$, while $a$ and $c$ are equal to the number of sites in $A$ and $C$. The state $\ket{I}$ is the maximally entangled state connecting $A\cup B$ and $C\cup D$, satisfying
\be
O_j^{C}\ket{I}=\left(O_j^{A}\right)^T\ket{I}\,,
\ee 
while $U_{AB}$ is the evolution operator acting non-trivially only on the system $A\cup B$. Then, using the orthogonality of the Pauli operators, and after simple simplifications, we have
\bea
{\rm tr}\left[\rho_{AC}^2\right]&=2^{-a-c-2N}\sum_{j,k}{\rm tr}\left[U^{\dagger}_{AB}O_j^AU_{AB}O_k^A\right]\nonumber\\
&\times {\rm tr}\left[O_k^A U^{\dagger}_{AB}O_j^AU_{AB}\right]\,.
\label{eq:1_spin}
\eea
Consider now the sum
\begin{align}
2^{-a-c-3N}\sum_{j,k,\tilde{j}, \tilde{k}}{\rm tr} \left[O^A_{\tilde{j}}O^B_{\tilde{k}}\left(U^{\dagger}_{AB}O_j^AU_{AB}O_k^A\right)O^A_{\tilde{j}}O^B_{\tilde{k}}\right.\nonumber\\
\times \left. \left(O_k^A U^{\dagger}_{AB}O_j^AU_{AB}\right)\right]\,.
\label{eq:2_spin}
\end{align}
Using the identity \cite{HQRY16}
\be
\sum_{j} A_{j} \mathcal{O} A_{j}=|A| \operatorname{tr}_{A}\{\mathcal{O}\}\,,
\label{eq:identity_spins}
\ee
(here $A_j$ are a complete basis of operators for the Hilbert space associated with $A$, while $|A|$ is its dimension), one immediately obtains that the r.h.s. of Eq.~\eqref{eq:1_spin} is equal to \eqref{eq:2_spin}. Therefore
\begin{align}
		{\rm tr}\left[\rho^2_{AC}\right]=&2^{-a-c-3N}\sum_{j,k,\tilde{j}, \tilde{k}}{\rm tr} \left[O^A_{\tilde{j}}O^B_{\tilde{k}}\left(U^{\dagger}_{AB}O_j^AU_{AB}O_k^A\right)\right.\nonumber\\
		\times& \left.O^A_{\tilde{j}}O^B_{\tilde{k}}\left(O_k^A U^{\dagger}_{AB}O_j^AU_{AB}\right)\right]\nonumber\\
		=&\frac{1}{2^{3N-a+c}}\sum_{j,l}{\rm tr} \left[O^B_l O^A_j(t)O^B _l O^A_j(t)\right]
\end{align}
In the last step, we summed over $k$, used once again the identity \eqref{eq:identity_spins}, and finally renamed the indexes $\tilde{k}=l$. Putting all together, we find
\be
\frac{1}{4^{a+d}}\frac{1}{2^N}\sum_{j,k}{\rm tr}\left[O^A_j(t)O^D_k(0)O^A_j(t)O^D_k(0)\right]=2^{N-a-d-S^{(2)}_{AC}}\,.
\ee
This is exactly the same result as in \cite{HQRY16}. An analogous derivation holds for the case of $S^{(2)}_{AD}$. This equation encodes a close connection between the tripartite information and the OTOCs, and allows one to establish that chaos, as measure by small values of all OTOCs, implies scrambling~\cite{HQRY16}. In the next section we show that a similar relation, with the addition of proper signs, holds in the case of fermionic systems.

\subsection{The case of Majorana operators}
\label{sec:app_majorana_case}

For Majorana operators one needs a different treatment. Indeed, the identity \eqref{eq:identity_spins} is no longer valid, but should be modified as follows. Let $\mathcal{O}$ be an operator with a well defined parity of Majorana operators, i.e. $\mathcal{O}$ is the sum of strings of operators that are either all even or all odd. Then, by expanding in the operator basis of Majorana operators, one can prove
\be
\sum_{j} (-1)^{\ell_j \ell_{\mathcal{O}}}A_{j} \mathcal{O} A^{\dagger}_{j}=|A| \operatorname{tr}_{A}\{\mathcal{O}\}.
\label{eq:identity_majorana}
\ee
Here $\ell_j$ is the length of the operator $A_j$. For example, if $A_j=\psi_1\psi_4$ then $\ell_j=2$. Analogously, $\ell_{\mathcal{O}}$ is the length of one of the terms in $\mathcal{O}$. Since all these terms have the same parity, it does not matter which one we choose. We can then proceed as in the previous sections, now paying attention to the order of the operators involved in the calculations. First, we have
\be
\rho_{AC}=\frac{1}{2^{a+c}}\sum_{j,k}\left(O_j^{A}O_k^{C}\right)^{\dagger}\braket{I|U^{\dagger}_{AB}O_j^{A}O_k^{C}U_{AB}|I}\,,
\ee
where now $a$ and $c$ are \emph{half} the number of sites in $A$ and $C$. Proceeding as before, we have
\begin{align}
{\rm tr}\left[\rho_{AC}^2\right]=2^{-a-c-2n}\sum_{j,k}{\rm tr}\left[U^{\dagger}_{AB}O_j^AU_{AB}\left(O_k^A\right)^{\dagger}\right]\nonumber\\
\times {\rm tr}\left[O_k^A U^{\dagger}_{AB}\left(O_j^A\right)^{\dagger}U_{AB}\right]\,,
\label{eq:1_majorana}
\end{align}
where $n=N/2$. Consider now the sum
\begin{align}
2^{-a-c-3n}\sum_{j,k,\tilde{j}, \tilde{k}}(-1)^{\ell_{\tilde{j}}(\ell_{j}+\ell_k)+\ell_{\tilde{k}}(\ell_{j}+\ell_k)}\nonumber\\
{\rm tr} \left[O^A_{\tilde{j}}O^B_{\tilde{k}}\left(U^{\dagger}_{AB}O_j^AU_{AB}\left(O_k^A\right)^{\dagger}\right)\left(O^B_{\tilde{k}}\right)^{\dagger}\left(O^A_{\tilde{j}}\right)^{\dagger}\right.\nonumber\\
\left.\left(O_k^A U^{\dagger}_{AB}\left(O_j^A\right)^{\dagger}U_{AB}\right)\right]\,.
\label{eq:2_majorana}
\end{align}
Noticing now that the evolution operator can always be written as sum of even strings of Majorana operators, one can directly apply the identity \eqref{eq:identity_majorana} to prove that the r.h.s. of \eqref{eq:1_majorana} is equal to \eqref{eq:2_majorana}. On the other hand, using
\be
\ell_{\tilde{j}}(\ell_{j}+\ell_k)+\ell_{\tilde{k}}(\ell_{j}+\ell_k)=\ell_{j}(\ell_{\tilde{j}}+\ell_{\tilde{k}})+\ell_{k}(\ell_{\tilde{j}}+\ell_{\tilde{k}})\,,
\ee
we can sum over $k$ by employing once again the identity \eqref{eq:identity_majorana}, and finally rename $\tilde{k}=r$. Putting all together, we obtain
\begin{align}
{\rm tr}\left[\rho^2_{AC}\right]=&\frac{1}{2^{3n-a+c}}  \sum_{j,r}(-1)^{\ell_j\ell_r}\nonumber\\
&\times {\rm tr} \left[O^B_r O^A_j(t)\left(O^B _r\right)^{\dagger}\left( O^A_j(t)\right)^{\dagger}\right]\,.
\end{align}
We see that additional signs appear in the sum over the OTOCs with respect to the case of Pauli matrices. However, the R\'enyi-$2$ entropy still encodes global information about the sum of the OTOCs over extended regions of the system.

\section{Details on the numerical implementation}
\label{sec:numerical_details}
In this section, we explain a few details for the numerical computation. For the OTOCs, we implement \eqref{eq:overlap_evolution} in terms of the modes $b_1,b_2,b_3,b_4$, as explained in the main text. For the entropies however, we use different modes. While we could likewise implement \eqref{eq:w_sac} and \eqref{eq:w_sad} in $b$-modes, this introduces large numerical error as $N$ increases, due to cancellations of large numbers. Instead we have found that for the entropy $S_{AC}^{(2)}$, a numerical calculation in terms of modes $c_1,c_2,c_3,c_4$ \eqref{eq:c-modes} is most stable. For the entropy $S_{AD}^{(2)}$, we found that using the modes $b_1,b_2,c_3,c_4$ gives the most stable results. For our numerical calculation we have therefore transformed initial state $\ket{\overline{\mathcal{U}}(0)}\rangle$ \eqref{eq:initial_state_bosonic}, effective Hamiltonian $H$ \eqref{eq:otoc hamiltonian} and $\ket{W_{S_{AC,BD}^{(2)}(\lbar)}}\rangle$ \eqref{eq:vec_AC}-\eqref{eq:vec_AD} into these modes.

\section{The R\'enyi-$2$ entanglement entropy~$S^{(2)}_{AD}(\lbar)$}
\label{sec:renyi_ad}

In this appendix,  we turn to the task of deriving the vector $|W_{S^{(2)}_{AD}(\bar\ell)}\rangle\rangle$ introduced in Eq.~\eqref{eq:w_sad}, corresponding to the exponential of the second R\'enyi entropy $S^{(2)}_{AD}(\lbar)$. The discussion goes along the same lines of the one presented in Sec.~\ref{sec:renyi_ac} for the entropy $S^{(2)}_{AC}(\lbar)$. Writing out the partial trace as done for the other entropy, we get
\begin{equation}
ρ_{AD} = \sum_{F^a_A,F^b_{D}} \frac{1}{2^{N/2}} (F^a_AF^b_{D})^\dag \braket{I^{ab}|U^{a\dag}(F_A^aF_{D}^b)U^a|I^{ab}}\,,
\end{equation}
where we denoted by $\{F^a_A\}$ and $\{F^b_D\}$ a complete basis for the operators in $A$, $D$ respectively; namely $F^a_A$ and $F^b_D$  take value in all the possible strings of Majorana operators supported in $A$ and $D$.
We can continue along the same lines as for the other entropy, giving
\begin{widetext}
\begin{gather}
\tr \left[ \overline{ρ_{AD}^2} \right] = \frac{1}{2^{N/2}\sqrt{N!n_1!\cdots n_{abcd}!}} \sum_{\vec{n}} c_{\vec{n}}(t) \sum_{\pi\in S_N} π (-1)^{\gamma_A+γ_B+δ}\, (*)\pi^{-1}\,.
 \label{eq:crossentropy insertion point}
\end{gather}
Here the term $(*)$ can be evaluated as for the other entropy up until (\ref{eq:crossentropy same up to here}).
	Then, however, $Ψ_{bB}^{a\dag}$ and $Ψ_{aB}^a$ are not the only parts in the first expression acting on this subspace, now $F_{D}^a$ also does. So, continuing from $(*)$ and dropping the doubled system label as all operators are in the same system, we have
	\begin{align}
	(*) &= \sum_{F_A,F_D} \braket{I^{ab}| Ψ_{bA}^{a\dag} Ψ_{bB}^{a\dag} F_A^a Ψ_{aB}^aΨ_{aA}^aF_D^{a\dag}|I^{ab}}\braket{I^{cd}|F_D^cΨ_{dA}^{c\dag}Ψ_{dB}^{c\dag} F_A^{c\dag} Ψ_{cB}^c Ψ_{cA}^c |I^{cd}} \nonumber\\
	&= \sum_{F_A,F_D} \braket{I| Ψ_{bA}^{\dag} F_A Ψ_{aA} Ψ_{bB}^\dag Ψ_{aB} F_D^\dag|I}\braket{I|F_D Ψ_{dB}^{\dag}Ψ_{cB}Ψ_{dA}^{\dag} F_A^{\dag} Ψ_{cA} |I}\,.
	\end{align}
	The left side is only non-zero for
	\begin{gather}
	Ψ_{bA}^{\dag} F_A Ψ_{aA} = \pm 1 \Rightarrow F_A^\dag = \pm Ψ_{aA}Ψ_{bA}^\dag \\
	Ψ_{bB}^\dag Ψ_{aB} F_D^\dag = \pm 1 \Rightarrow F_D = \pm  Ψ_{bB}^\dag Ψ_{aB}
	\end{gather}
	such that we can evaluate the sum $\sum_{F_A,F_{D}}$, inserting these in the right side. We get
	\begin{align}
	(*) &= \braket{I|I} \braket{I|Ψ_{bB}^\dagΨ_{aB}Ψ_{dB}^\dagΨ_{cB}\, Ψ_{dA}^\dag Ψ_{aA}Ψ_{bA}^\dag Ψ_{cA}|I} \nonumber\\ 
	&= \tr Ψ_{bB}^\dag Ψ_{aB}Ψ_{dB}^\dag Ψ_{cB} / 2^{\lbar/2}\, \tr Ψ_{dA}^\dag Ψ_{aA}Ψ_{bA}^\dag Ψ_{cA} / 2^{\lbar/2}\nonumber \\
	&= \tr Ψ_{dB}Ψ_{cB}Ψ_{bB}Ψ_{aB} / 2^{\lbar/2}\, \tr Ψ_{aA}Ψ_{bA}Ψ_{cA}Ψ_{dA}/2^{\lbar/2} (-1)^{\frac{n_b+n_d}{2} + n_{bB} + n_{dB}}\nonumber\\
	&= \tr (Ψ_{aB}Ψ_{bB}Ψ_{cB}Ψ_{dB})^\dag / 2^{\lbar/2}\, \tr Ψ_{aA}Ψ_{bA}Ψ_{cA}Ψ_{dA}/2^{\lbar/2} (-1)^{\frac{n_b+n_d}{2} + n_{bB} + n_{dB}}\nonumber\\
	&\qquad (-1)^{(n_{aB}(n_{aB}-1)+n_{bB}(n_{bB}-1)+n_{cB}(n_{cB}-1)+n_{dB}(n_{dB}-1))/2} \nonumber\\
	&= (-1)^{γ_A+γ_{B}} (-1)^{(n_b+n_d)/2} (-1)^{(n_{aB}(n_{aB}-1)+n_{bB}(n_{bB}+1)+n_{cB}(n_{cB}-1)+n_{dB}(n_{dB}+1))/2}\,.
	\end{align}
	The traces now give $(-1)^{γ_A}$ and $(-1)^{γ_B}$, respectively. Inserted back into (\ref{eq:crossentropy same up to here}), these cancel, and the $δ$ partially cancels the other phases,
	\begin{align}
	\tr \left[\overline{ρ^2_{AD}} \right]= \frac{1}{2^{N/2}\sqrt{N!n_1!\cdots n_{abcd}!}}\sum_{\vec n}c_{\vec n}(t) \sum_{π\in S_N}π (-1)^{(n_b+n_d)/2}(-1)^{(-n_{aB}  + n_{bB} - n_{cB} + n_{dB})/2} π^{-1}\,.
	\end{align}
Again, this has the form \eqref{eq: W_N} with suitable choices of $\alpha$'s. Then, according to Eq.~\eqref{eq: W_N result} we obtain
\bea
|W_{S^{(2)}_{AC}(\bar\ell)}\rangle\rangle=\frac{1}{2^{N/2}\sqrt{N!}}\left[a_{1}^{\dagger}+ia_{ab}^{\dagger}+ia_{ad}^{\dagger}+ia_{bc}^{\dagger}+ia_{cd}^{\dagger}-a_{abcd}^{\dagger}+a_{ac}^{\dagger}-a_{bd}^{\dagger}\right]^{\ell}\nonumber\\ \times \left[
a_{1}^{\dagger}+ia_{ab}^{\dagger}+ia_{ad}^{\dagger}+ia_{bc}^{\dagger}+ia_{cd}^{\dagger}-a_{abcd}^{\dagger}-a_{ac}^{\dagger}+a_{bd}^{\dagger}
\right]^{\bar\ell}\ket{\Omega}\,.
\eea
Finally, the transformation to $b$-modes in \eqref{eq:b_modes} results in Eq.~\eqref{eq:vec_AD}.
\end{widetext}

\end{document}